\documentclass[twoside,twocolumn,9pt]{article}
\usepackage{extsizes}
\usepackage[super,sort&compress,comma]{natbib} 
\usepackage[version=3]{mhchem}
\usepackage[left=1.5cm, right=1.5cm, top=1.785cm, bottom=2.0cm]{geometry}
\usepackage{balance}
\usepackage{times,mathptmx}
\usepackage{sectsty}
\usepackage{graphicx} 
\usepackage{lastpage}
\usepackage[format=plain,justification=raggedright,singlelinecheck=false,font={stretch=1.125,small,sf},labelfont=bf,labelsep=space]{caption}
\usepackage{float}
\usepackage{fancyhdr}
\usepackage{fnpos}
\usepackage[english]{babel}
\usepackage{array}
\usepackage{droidsans}
\usepackage{charter}
\usepackage[T1]{fontenc}
\usepackage[usenames,dvipsnames]{xcolor}
\usepackage{setspace}
\usepackage[compact]{titlesec}
\usepackage{multirow}
\usepackage{subfigure}
\usepackage{cancel}
\newcommand{\rcom}[1]{\textcolor{black}{\textsf{{#1}}}}

\usepackage{epstopdf}%This line makes .eps figures into .pdf - please comment out if not required.

\definecolor{cream}{RGB}{222,217,201}

\begin{document}

\pagestyle{fancy}
\thispagestyle{plain}
\fancypagestyle{plain}{

%%%HEADER%%%
\fancyhead[C]{\includegraphics[width=18.5cm]{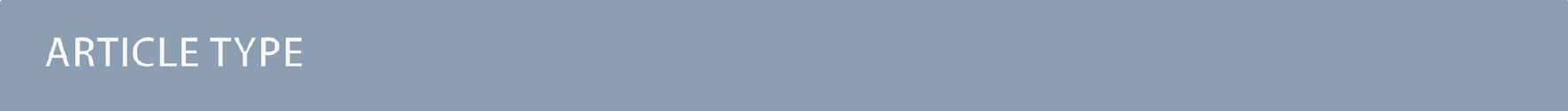}}
\fancyhead[L]{\hspace{0cm}\vspace{1.5cm}\includegraphics[height=30pt]{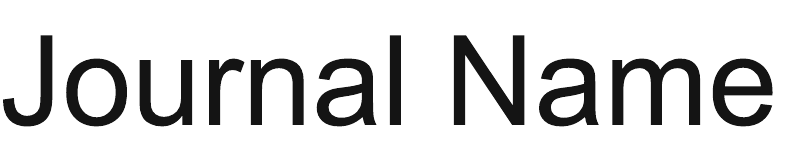}}
\fancyhead[R]{\hspace{0cm}\vspace{1.7cm}\includegraphics[height=55pt]{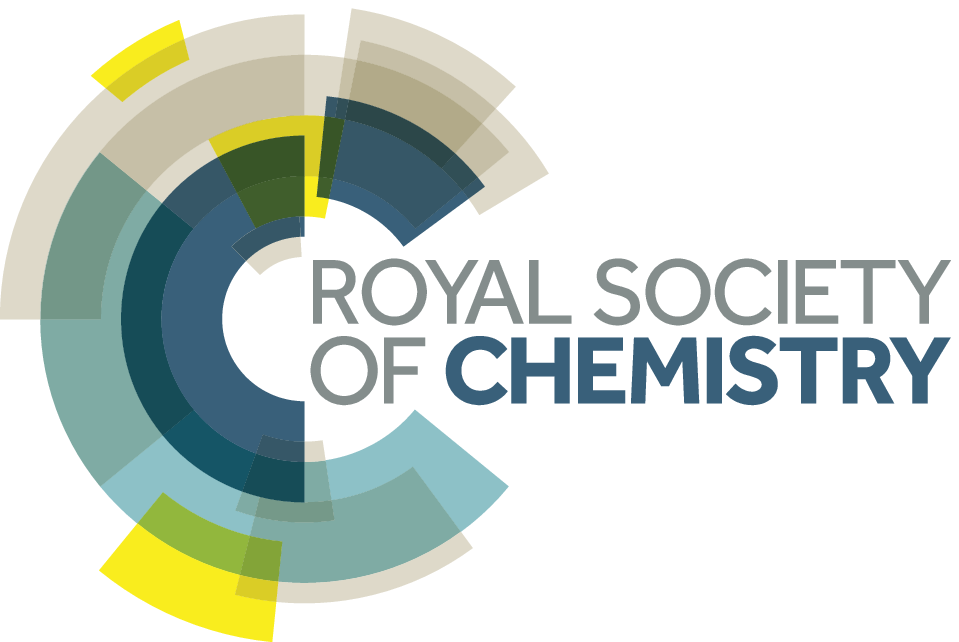}}
\renewcommand{\headrulewidth}{0pt}
}
%%%END OF HEADER%%%

%%%PAGE SETUP - Please do not change any commands within this section%%%
\makeFNbottom
\makeatletter
\renewcommand\LARGE{\@setfontsize\LARGE{15pt}{17}}
\renewcommand\Large{\@setfontsize\Large{12pt}{14}}
\renewcommand\large{\@setfontsize\large{10pt}{12}}
\renewcommand\footnotesize{\@setfontsize\footnotesize{7pt}{10}}
\makeatother

\renewcommand{\thefootnote}{\fnsymbol{footnote}}
\renewcommand\footnoterule{\vspace*{1pt}% 
\color{cream}\hrule width 3.5in height 0.4pt \color{black}\vspace*{5pt}} 
\setcounter{secnumdepth}{5}

\makeatletter 
\renewcommand\@biblabel[1]{#1}            
\renewcommand\@makefntext[1]% 
{\noindent\makebox[0pt][r]{\@thefnmark\,}#1}
\makeatother 
\renewcommand{\figurename}{\small{Fig.}~}
\sectionfont{\sffamily\Large}
\subsectionfont{\normalsize}
\subsubsectionfont{\bf}
\setstretch{1.125} %In particular, please do not alter this line.
\setlength{\skip\footins}{0.8cm}
\setlength{\footnotesep}{0.25cm}
\setlength{\jot}{10pt}
\titlespacing*{\section}{0pt}{4pt}{4pt}
\titlespacing*{\subsection}{0pt}{15pt}{1pt}
%%%END OF PAGE SETUP%%%

%%%FOOTER%%%
\fancyfoot{}
\fancyfoot[LO,RE]{\vspace{-7.1pt}\includegraphics[height=9pt]{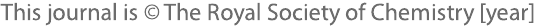}}
\fancyfoot[CO]{\vspace{-7.1pt}\hspace{13.2cm}\includegraphics{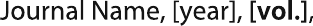}}
\fancyfoot[CE]{\vspace{-7.2pt}\hspace{-14.2cm}\includegraphics{head_foot/RF}}
\fancyfoot[RO]{\footnotesize{\sffamily{1--\pageref{LastPage} ~\textbar  \hspace{2pt}\thepage}}}
\fancyfoot[LE]{\footnotesize{\sffamily{\thepage~\textbar\hspace{3.45cm} 1--\pageref{LastPage}}}}
\fancyhead{}
\renewcommand{\headrulewidth}{0pt} 
\renewcommand{\footrulewidth}{0pt}
\setlength{\arrayrulewidth}{1pt}
\setlength{\columnsep}{6.5mm}
\setlength\bibsep{1pt}
%%%END OF FOOTER%%%

%%%FIGURE SETUP - please do not change any commands within this section%%%
\makeatletter 
\newlength{\figrulesep} 
\setlength{\figrulesep}{0.5\textfloatsep} 

\newcommand{\topfigrule}{\vspace*{-1pt}% 
\noindent{\color{cream}\rule[-\figrulesep]{\columnwidth}{1.5pt}} }

\newcommand{\botfigrule}{\vspace*{-2pt}% 
\noindent{\color{cream}\rule[\figrulesep]{\columnwidth}{1.5pt}} }

\newcommand{\dblfigrule}{\vspace*{-1pt}% 
\noindent{\color{cream}\rule[-\figrulesep]{\textwidth}{1.5pt}} }

\makeatother
%%%END OF FIGURE SETUP%%%

%%%TITLE, AUTHORS AND ABSTRACT%%%
\twocolumn[
  \begin{@twocolumnfalse}
\vspace{3cm}
\sffamily
\begin{tabular}{m{4.5cm} p{13.5cm} }

\includegraphics{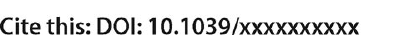} & \noindent\LARGE{\textbf{Modeling growth paths of interacting crack pairs in elastic media$^\dag$}} \\%Article title goes here instead of the text "This is the title"
\vspace{0.3cm} & \vspace{0.3cm} \\

 & \noindent\large{Ramin Ghelichi\textit{$^{a}$}  and Ken Kamrin$^{\ast}$\textit{$^{b}$}} \\%Author names go here instead of "Full name", etc.

\includegraphics{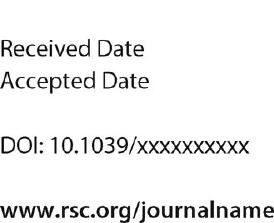} & \noindent\normalsize{The problem of predicting the growth of a system of cracks, each crack influencing the growth of the others, arises in multiple fields.  We develop an analytical framework toward this aim, which we apply to the `En-Passant' family of crack growth problems, in which a pair of initially parallel, offset cracks propagate nontrivially toward each other under far-field opening stress.  We utilize boundary integral and perturbation methods of linear elasticity,  Linear Elastic Fracture Mechanics, and common crack opening criteria to calculate the first analytical model for curved En-Passant crack paths.  The integral system is reduced under a hierarchy of approximations, producing three methods of increasing simplicity for computing crack paths.  The last such method is a major highlight of this work, using an asymptotic matching argument to predict crack paths based on superposition of simple, single-crack fields.  Within the corresponding limits of the three methods, all three are shown to agree with each other.  We provide comparisons to exact results  and existing experimental data to verify certain approximation steps.} \\%The abstrast goes here instead of the text "The abstract should be..."

\end{tabular}

 \end{@twocolumnfalse} \vspace{0.6cm}

  ]
%%%END OF TITLE, AUTHORS AND ABSTRACT%%%

%%%FONT SETUP - please do not change any commands within this section
\renewcommand*\rmdefault{bch}\normalfont\upshape
\rmfamily
\section*{}
\vspace{-1cm}

%%%FOOTNOTES%%%

\footnotetext{\textit{$^{a}$~Address, 77 Massachusetts Ave
Cambridge, MA, 02139, USA.}}
\footnotetext{\textit{$^{b}$~Address, 77 Massachusetts Ave
Cambridge, MA    02139, USA. Fax: (617) 258-8742; Tel: (617) 715-4157; E-mail: kkamrin@mit.edu}}
\footnotetext{\textit{$\ast$ Corresponding author}}

%%%END OF FOOTNOTES%%%

%\begin{keyword} Fracture mechanics, crack paths, crack interactions, analytical methods \end{keyword}

\section{Introduction}
\begin{figure*}[t!]
  \centering
  \includegraphics[width=0.8\textwidth]{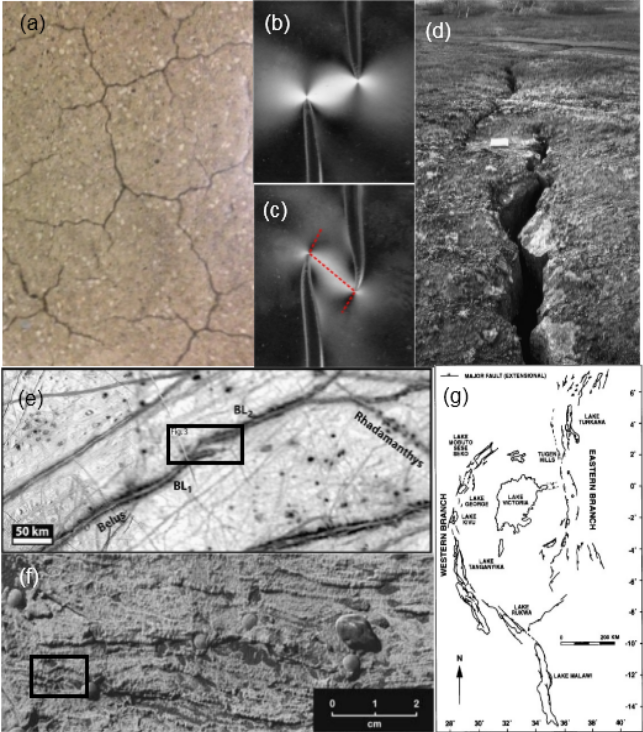}
  \caption{(a) A system of multiple cracks in flooring (b) Deviation stage of En-Passant cracks in soft polymeric material \cite{fender2010} (c) Two cracks move toward each other after initial divergence \cite{fender2010} (d) Example of interacting opening-mode extensional fractures at Krafla \cite{acocella2000} (e) Context image taken from USGS controlled photo mosaic of Europa (map I-2757) that includes portions of the prominent ridge complex Belus Linea (BL1 and BL2) and double ridge Rhadamanthys Linea \cite{bechtle2010} (f) Comparison of fine-scale morphologic features of bedrock from Meridiani Planum on Mars (Image by NASA/JPL and MER team) \cite{mccollom2005} (g) Index map of the East Africa rift system showing both the Eastern and Western branches \cite{nelson2000}} %(g) Parallel joints set in the dolomite layer of Argot stream, central Dead Sea basin \cite{sagy2001} (h) a photo from MIT pavement}
  \label{fig:EPexample}
  \vspace{-10pt}
\end{figure*}

The interaction of multiple cracks in brittle materials is a phenomenon observed across a variety of disciplines, from the study of human bones in biology \cite{koester2008}, to the mechanics of soft gels \cite{fender2010}, to the dynamics of planetary \rcom{crusts and tectonic plates} in geophysics \cite{vannucchi2008,acocella2000,bahat1984}, see Fig.\ref{fig:EPexample}.
Large-scale challenges arising from crack interactions include well-water contamination due to fracking \cite{osborn2011}, penetration of microbial life deep in the earth's crust \cite{pederson1997}, permeability of groundwater through naturally-occurring aquifers \cite{white2002}, and the geosequestration of CO$_2$ \cite{kelemen2008}. 
In these cases, the development of analytical and numerical methods for the growth dynamics of multiple cracks has been a decades-long challenge \cite{swain1974}.  More recently, the study of fracture patterns in soft materials has become a topic of specific interest due to the novel applications of these materials \cite{kundu2009, xu2013,lazarus2011,seitz2009,fender2010}.  %Particular attention has been paid to the fracture of gels and colloids.%; different experiments are performed to study the combined elasticity and fracture of polymer networks \cite{kundu2009}, rate dependence of fracture of copolymer gels \cite{seitz2009}, influence of film thickness of colloidal particles consolidates on the final crack morphologies \cite{lazarus2011}, effect of drying on the fracture behaviour of colloidal materials\cite{xu2013}, and the interaction of cracks in gels \cite{fender2010}. 
Particular attention has been paid to the formation and paths of cracks in gels and colloids.  Most of these studies have been experimentally driven, and include analyses of the crack patterns in drying colloids and the influence of film thickness  \cite{lazarus2011}, rate-dependent crack morphology and propagation patterns in copolymer gels \cite{seitz2009}, oscillatory fracture patterns in thin polymer sheets due to out-of-plane bending \cite{audoly2005}, and the curved growth paths of systems of multiple cracks in gels \cite{fender2010}. Accurate and general analytical models for fracture processes like these are in much need and carry fundamental importance, which has motivated our current study.  The growth paths of multiple cracks is a complex problem analytically because every increment of crack growth modifies the stress field globally and changes the stress intensity at all other cracks --- hence, it is a coupled, non-localized, inverse problem.

%Additionally, acquiring a better understanding of the mechanical behaviour of soft materials is crucial to the their growing applications; in this regard, recently, excessive attentions have been paid to the fracture of gels; different experiments are performed to study the elasticity and fracture behavior of swollen polymer networks \cite{kundu2009}, rate dependence of fracture of copolymer gels \cite{seitz2009}, influence of film thickness of colloidal particles consolidates on the final crack morphologies \cite{lazarus2011}, effect of drying on the fracture behavoir of colloidal materials\cite{xu2013}, and the interaction of cracks in gels \cite{fender2010}. 
%

Herein, we develop a novel toolset to predict kinked/curved crack paths in elastic media with multiple cracks, which will be applied to the model case of the ``En-Passant'' (EP) family of crack geometries. En-Passant cracks, named by Kranz \cite{kranz1979}, are two initially parallel offset cracks that grow under transverse loading and eventually approach each other through their propagation paths. Fig. \ref{fig:EPexample}(b) shows EP cracks in a soft gel; the growth paths of the cracks are initially \rcom{repelled} and then move toward each other, as seen in Fig. \ref{fig:EPexample}(c), due to lateral far-field tensile stress. Whether EP cracks initially \rcom{repel}  before approaching is an interesting notion in itself and depends on the crack placements; as lateral separation vanishes and the cracks becomes colinear, Melin \cite{melin1983} has shown that the crack paths \rcom{repel} and the straight-ahead path is unstable. EP crack phenomena was studied directly for the first time in the early 1970's \cite{lange1968, swain1978, yokobori1971}. EP cracks are reported repeatedly as one of the causes in the formation of ridges and crusts on the earth and other planets \cite{vannucchi2008, acocella2000,  bahat1984, mccollom2005, nelson2000, sempere1986, acocella2008,  patterson2010, niles2009}; some examples have been presented in Fig.\ref{fig:EPexample}(d)-(g).

In order to predict the growth paths in these systems, the first step is to determine a method for calculating stress fields in a body with multiple, arbitrary  cracks.
%The exact stress field is expressible in terms of a non-trivial system of integral equations over the body, which depend on the paths of all system cracks.
Different solution approaches for multi-crack stress fields have been presented in the last decades, which we briefly discuss now. For a system of two straight cracks in arbitrary positions, Isida \cite{isida1970} has presented an analytical method based on Laurent series expansions for the stress field; the solution, which can be considered one of the most accurate ones, ends with a linear system of equations yielding the coefficients of the Laurent series. Yokobori \cite{yokobori1971} has used a continuous distribution of infinitesimal dislocations to calculate stress intensity factors for offset straight cracks, which has been used to approximate EP kink angles. Savruk \cite{savruk1975}, also utilizing the dislocation-density-based formulation, produced a set of integral equations that can be solved numerically to calculate the stress field in a system of multiple cracks of arbitrary shape. His work has been extended by Chen \cite{ChenBook2003}, who has proposed algorithms for a variety of integral equations.   
Hori and Nemat-Nasser \cite{hori1985} have reduced the stress calculation for arbitrarily located pairs of straight cracks to a linear system of equations by using a Taylor series with unknown constants. Kachanov \cite{Kachanov1987}, by a simple ``alternative method", estimates stress intensity factors for a system of straight cracks by canceling the residual mean traction from the known solution of a single crack. Many other efforts have been made to solve or approximate the integral equations of Muskhelishvili's method for the stress field, including work done by Ukadgaonker and Naik \cite{ukadgaonker1991} in a series of articles which contain various solutions for interacting cracks. The Schwarz alternating method \cite{sobolev1939}  has been utilized in order to reduce ``multiple connected regions'' in a two-crack problem to a sequence of simply connected regions. The Sih method \cite{sih1973} can then be used to find the crack propagation angle in the first step of the opening. 
Herein we develop and verify a sequence of analytical models for the propagation paths of EP cracks, achieved under the assumptions of Linear Elastic Fracture Mechanics.  

Our solution approach is rooted in perturbation methods applied to the Muskhelishvili formulation for elastic stress fields \cite{muskhelishvili1952}. The procedure we discuss is therefore applicable in plane stress or plane strain conditions. Three different formulae will be presented for the growth paths in a general system of EP cracks, each formula simpler than the previous at the expense of certain additional approximation errors.  All three methods use the \emph{local symmetry} criterion for mixed-mode crack opening.  
%%%%%
%%%%%
Our first method is obtained by simultaneously solving the system of integral equations plus local symmetry to approximate the growth paths of both cracks.  By expanding the integral system in series and applying perturbation methods, we obtain a non-linear system of equations, whose solution yields the crack extension paths along with the stress field after the extension.  Our second method uses the stress field of the initial pair of parallel offset cracks and considers the growth of one crack without considering the growth of the opposing crack, an assumption valid for small extension paths. We verify these two methods against an exact solution for the stress field. The third method, which is a highlight of this paper, is far simpler than either of the two prior ones.  It uses a matched-asymptotic-expansion argument, of the type commonly utilized in fluid mechanics problems (e.g. \cite{holmes2013}); matching techniques are less common in solid mechanics though have had some recent use \cite{leguillon2011}.  Herein, we conjoin an ``inner solution'' for stress near the crack tip --- a Williams expansion \cite{williams1952} with undetermined coefficients --- to a simple ``outer solution'' for the stress-field in a region not close to either crack  --- obtained by superimposing solutions for two isolated cracks.  Upon matching the fields, approximate Williams expansion coefficients are obtained, which can be substituted directly into the closed formula of the second method to predict paths.  While the first two methods are useful in their own right, they also play an important role in the current work to verify the accuracy of the much-simpler third method in a variety of EP geometries and loading conditions.  
\section{Analytical solution procedure}
\begin{figure}[t]
  \includegraphics[width=0.5\textwidth]{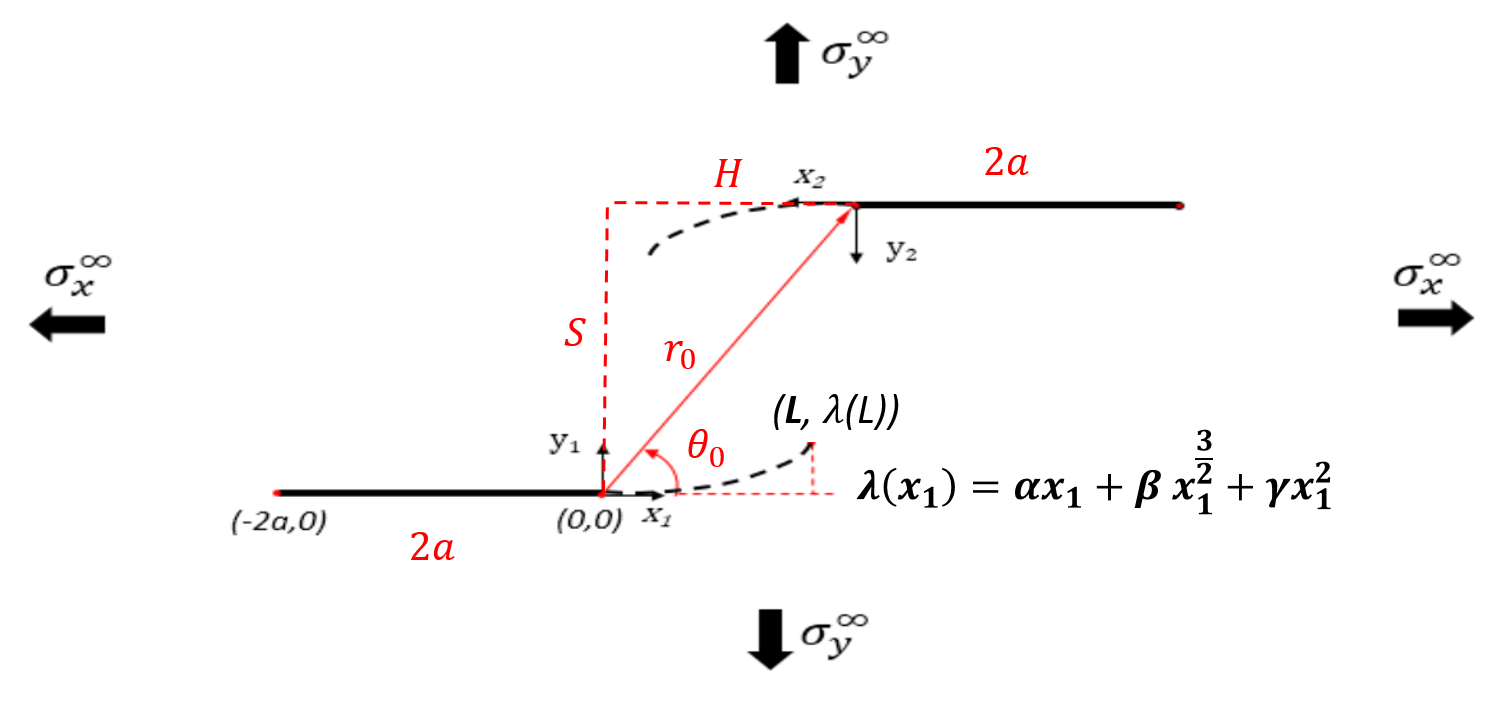}
  \caption{A system of EP cracks, which grow in a curvilinear path $\lambda(x)$ due to far-field loading. \rcom{In this image, $2a$ is the length of each crack,  and $S=r_0 \sin\theta_0$ and $H=r_0 \cos\theta_0$ are the vertical and horizontal distances of two crack tips respectively}. 
}
     \label{fig:crack}
\end{figure}
\indent Experimental results and numerical models of EP cracks \cite{swain1974, fender2010,swain1978, ukadgaonker1991, segall1980, melin1983, bazant1986, kamaya2010} confirm that cracks generally follow a curvilinear path. 
Therefore, in view of Fig. \ref{fig:crack}, we assume the propagation path is a curve $\lambda(x)$ extending the lower-left crack, with a symmetric path for the other.  The cracks are initially straight non-coplanar parallel cracks with a distance of $r_0$ between their tips and an angle of $\theta_0$, $\pi/2 <\theta_0<0$, from the horizontal to the line connecting their tips.  The material is deemed homogeneous, isotropic, and linearly elastic with brittle fracture behavior. 

First, we outline the strategy for computing the stress field in the system when the cracks have opened along some given path $\lambda(x)$. Later, this relation will be solved simultaneously with the opening criterion to determine $\lambda$. Our approach, as indicated in Fig.\ref{fig:superposition}, is to construct the stress field as a superposition of three different fields, denoted $A$, $B$, and $C$:
\begin{figure*}[t]
\includegraphics[width=0.9\textwidth]{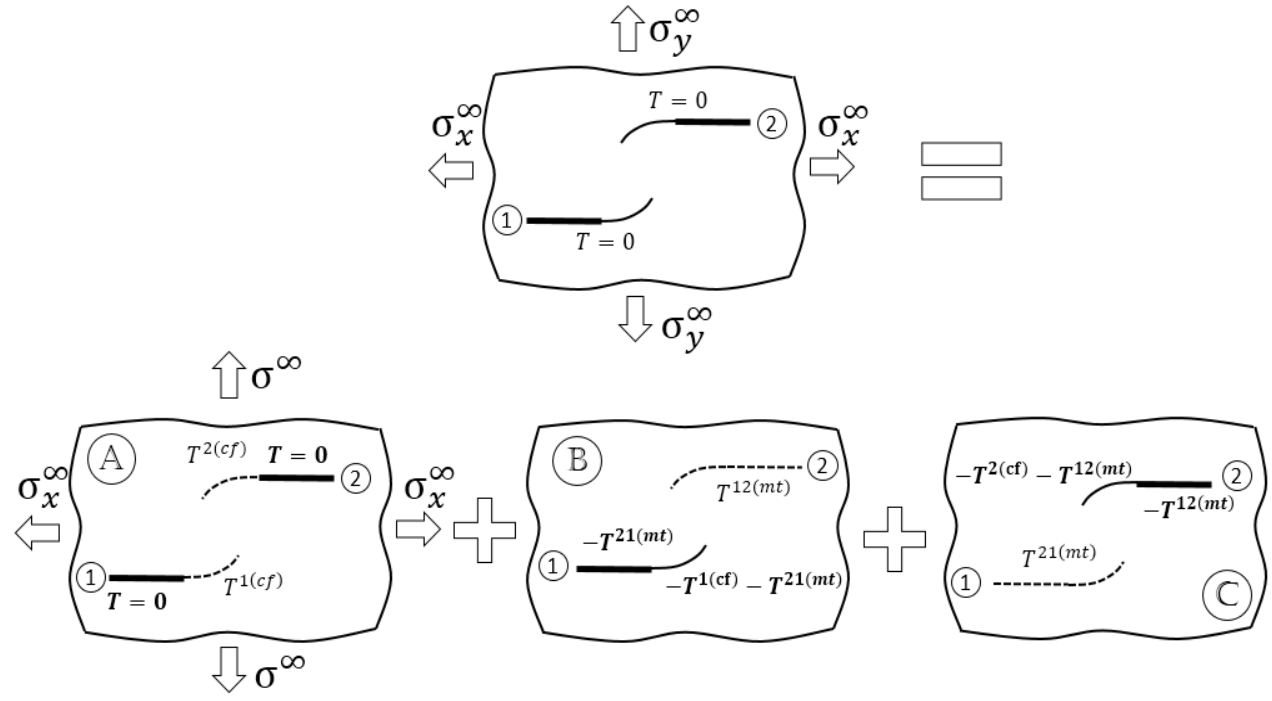}
\caption{Superposition of different elasticity solutions (Problems A, B, and C) to represent stresses in a system of extended, curved cracks. Tractions are assigned on solid lines as boundary conditions to produce a stress field.  The thicker lines indicate the original cracks before growth.  Other tractions needed in the calculation are evaluated on the dashed lines.
}
\label{fig:superposition}
\end{figure*}
\begin{itemize}
\item Problem $A$ is the solution for two offset straight parallel cracks under biaxial loading; this is assumed to be the initial state of the cracks before extension.  The stress solution has been proposed by Hori and Nemmat-Nasser \cite{hori1985} or Isida \cite{isida1970}.  The solution is characterized by a mixed-mode loading on both cracks, due to the crack-crack interaction \cite{lange1968, yokobori1971, ukadgaonker1991, isida1970, savruk1975, hori1985}.  The main goal in this step is to determine the ``close-field'' tractions, $T^{1(cf)}$ and $T^{2(cf)}$, on the extension path by resolving the calculated stress field from Problem $A$ along $\lambda(x)$. \\

\item Problems $B$ and $C$ relate to the modification of Problem A needed to represent extended curvilinear cracks. We assign tractions on the extended cracks as follows.  Part of the traction cancels stress from Problem $A$, i.e. we apply  $T^{1(cf)}$ and $T^{2(cf)}$ on the extensions in $B$ and $C$ respectively. In addition to the traction coming from Problem $A$, Problems $B$ and $C$ together form a closed inverse problem for the remaining ``mirror tractions.'' 
That is, Problem $B$ requires the mirror traction $T^{21(mt)}$ as a boundary condition, which is given from the solution of Problem $C$, but Problem $C$ requires the inputting of a mirror traction given by the solution for Problem $B$. To solve the joint problem, we use boundary integral equations \cite{muskhelishvili1952} to express the stress fields in Problems $B$ and $C$ in terms of the unknown tractions ($T^{12(mt)}$ and $T^{21(mt)}$) and the extension path.  This integral method gives (planar) stress fields in terms of two biharmonic complex functions. These functions can be approximated for any path $\lambda(x)$ using perturbation methods in the limit of small deflections \cite{CotterellRice1980, sumi1983}.  We then use the symmetry of the crack extensions to reduce to a single unknown mirror traction, which is then solved using a Taylor series.
%The tractions can be replaced as a sum of the asymptotsolution \cite{williams1952,williams1957} of the existing field for two offset straight parallel cracks \cite{yokobori1971,isida1970,hori1985} ($T_1^{cf}$ and $T_2^{cf}$ in Fig.\ref{fig:superposition}) obtained from Problem $A$ which is justified by the perturbation method \cite{CotterellRice1980, karihaloo1981, sumi1983}; and second, the``mirror traction" which comes from the effect of the other crack's extension on the stress field; the mirror tractions are shown by $T_1^{ms}$ and $T_2^{ms}$ in Fig.\ref{fig:superposition}. After the latter replacement 
The details of the solution procedure are in Section \ref{sec:curvedcrack}.  
\end{itemize}
The stress field is then a function of the crack extension path $\lambda(x)$. Note that we only model extensions of the inner crack tips;  these have the higher mode-I stress intensity factor and will open first in the limit of a stiff elastic response.  This assumption is further validated if we restrain to small crack extensions.  Moreover, as the cracks extend, we assume quasi-static crack growth.  That is to say, we assume the ratio $\sigma_x^{\infty}/\sigma_y^{\infty}$ is fixed but we let the magnitudes of $\sigma_{\infty}^x$ and $\sigma_{\infty}^y$ arise by the condition that the cracks remain critically loaded in opening during growth.
According to the local symmetry criterion \cite{banichuk1970, goldstein1974} and many other models for crack propagation in brittle homogeneous isotropic material  \cite{hussain1927,sih1991}, the cracks propagate in a path in which the tip is in Mode I condition; i.e. if $K_{II}\neq0$ at the crack tip, the crack first kinks and then opens in a path through which it can maintain $K_{II}=0$ \cite{CotterellRice1980}.  With the aforementioned method for generating the stress field and corresponding stress intensity factors, the crack path $\lambda(x)$ is identified by requiring that $K_{II}=0$ as the cracks grow along this path. 
\section{Crack path calculation}
\label{sec:curvedcrack}
In plane-stress conditions, linear elastic stress fields can be expressed by two biharmonic complex functions of $z=x+iy$, the Muskhelishvili potentials \cite{muskhelishvili1952}:
\begin{equation}
\begin{aligned}
& \sigma_{xx}+\sigma_{yy} = 2[\phi(z)+\overline{\phi(z)}] \\
& \sigma_{yy}-\sigma_{xx} + 2 i \sigma_{xy} = 2[(z-\bar{z})\overline{\phi'(z)}+\Omega(\bar{z})-\overline{\phi(z)}]. \\
\end{aligned}
\label{eq:mus}
\end{equation}
In a system of EP cracks, we express $\phi$ and $\Omega$ for the superposition of Problems $B$ and $C$, as a sum of two such potentials, one representing the solution for Problem $B$ (denoted from here on with superscript $1$) and the other for Problem $C$ (denoted with superscript $2$).  That is,
\begin{equation}
\begin{aligned}
& \phi(z)=\phi^1(z_1)+\phi^2(z_2)\\
& \Omega(z)=\Omega^1(z_1)+\Omega^2(z_2)\\
& z=z_1=x_1+iy_1, \ z_2=x_2+iy_2=z_0-z_1, \ z_0=r_0 e^{i\theta_0}\\
\end{aligned}
\end{equation}
Figure \ref{fig:crack} shows the relation between the $z_1$ and $z_2$ coordinate systems. We assume symmetry consistent with Fig. \ref{fig:crack} such that the path for both cracks is given by the same function $\lambda$, i.e. crack 1 follows $y_1=\lambda(x_1)$ and crack 2 follows $y_2=\lambda(x_2)$. Each of the $\phi^j$ and $\Omega^j$ for $j=1$ or $2$ can be approximated up to the first order in $\lambda$ (for higher order analysis, see \cite{karihaloo1981}) by the equations
\begin{equation}
\begin{aligned}
& \phi^j(z_j) = \phi^j_0(z_j) + \phi^j_1(z_j) + O(\lambda^2) \\
&  \Omega^j(z_j) = \Omega^j_0(z_j) + \Omega^j_1(z_j) + O(\lambda^2) \\
\end{aligned}
\label{eq:phi0}
\end{equation}
where $\phi^j_0$ and $\Omega^j_0$ are solutions for two straight crack extensions, and $\phi^j_1$, and $\Omega^j_1$ adjust for crack path deviations  \cite{CotterellRice1980,sumi1983}. Both can be expressed as integrals of the traction on crack surfaces as shown below \cite{sumi1983}
\begin{equation}
\begin{aligned}
&  \phi ^j_0(z_j) = \Omega ^j_0(z_j) = \frac{1}{2\pi \sqrt{z_j-L}}\int_{0}^L(T^j_n(t)-i T^j_s(t)) \frac{\sqrt{L-t}}{z_j-t}dt \\
& \phi ^j_{1}(z_j) + \Omega^j_{1}(z_j) = \\
& \ \frac{1}{\pi \sqrt{z_j-L}} \int_{0}^L (\eta(t)  T'^j_s(t)+2\eta' (t)T^j_s(t)-i \eta(t) T'^j_n(t))\frac{\sqrt{L-t}}{z_j-t}dt.
\end{aligned}
\label{eq:fandw}
\end{equation}
We define $\eta(t)=\lambda(t)-\lambda(L)$. 
$T^j_n(t)$ and $T^j_s(t)$ as the normal and shear tractions at location $x_j=t$ as prescribed on the crack for each of Problem B and C. 
As shown in Fig.\ref{fig:superposition}, the tractions on the crack extension can be expressed as the superposition of two different traction distributions: a \emph{close-field traction}, presumably known from the initial crack geometry, and an unknown \emph{mirror traction}.  That is, for $0\le t\le L$,
\begin{equation}
\begin{aligned}
%&\text{for $0\le t\le L$.:} \\
&T^j_n(t)-i T^j_s(t)=T_n^{kj(mt)}(t)-i T_s^{kj(mt)}(t)+T_n^{j(cf)}(t)-i T_s^{j(cf)}(t) \\ 
&k=1,2; \ \ k  \neq j.
\end{aligned}
\label{eq:traction1}
\end{equation}
In the integrals above, as an approximation that we will validate in a moment, we have assumed we can neglect residual mirror tractions on the original cracks, as long as the cracks begin far enough apart.  This allows us to make the simplification
\begin{equation}
T^{kj(mt)}_n(t)-i T^{kj(mt)}_s(t)\approx 0 \ \ \ \text{for $-2a\le t \le 0; \ \ k=1,2; \ \ k  \neq j$}.
\label{eq:traction}
\end{equation}
Symmetry of the geometry causes symmetric extensions, which implies:  
\begin{equation}
T^{12(mt)}(t)=T^{21(mt)}(t)\equiv T^{(mt)}(t) \ \ \text{for} \ \ 0\le t\le L
\end{equation}
Similarly, geometric symmetry requires that 
\begin{equation}\label{moresym}
\phi^1(z)=\phi^2(z) \ \ \text{and} \ \ \Omega^1(z)=\Omega^2(z).
\end{equation} In view of Fig. \ref{fig:superposition} Problem B, the stress field in Problem $B$ evaluated at the location of crack 2 defines the mirror traction $T^{12(mt)}$.  Therefore,
\begin{equation}\label{eq:system}
\begin{aligned}
&-(T_n^{12(mt)}(x_2)-i T_s^{12(mt)}(x_2))= \phi^1_0(z_0-x_2)+\Omega^1_0(z_0-x_2)+\\
& \hspace{1cm}+\phi^1_1(z_0-x_2)+\Omega^1_1(z_0-x_2)+\\ 
& \hspace{1cm}+i\eta(x_2)[\phi^1_0(z_0-x_2)+\Omega^1_0(z_0-x_2)]'+\\
& \hspace{1cm} +2i[\eta(x_2)( \overline{\phi^1_0(z_0-x_2)}-\Omega^1_0(z_0-x_2))]'. %\right|_{z=z_0-x_1} 
\end{aligned}
\end{equation} 
Upon substituting Eqs \ref{eq:fandw}-\ref{moresym} into the above, we obtain a closed integral equation for $T^{(mt)}$.  Solving this equation is a key step in the work of this paper.

The relationship between the stress intensity factors at the inner crack tips and the biharmonic functions ($\phi^j$ and $\Omega^j$) are presented in Eq.\ref{eq:integ}
\begin{equation}
\begin{aligned}
K^j_I(L)-i K^j_{II}(L)&=\lim_{r_j\rightarrow0}\sqrt{2\pi r_j}\bigg[2\phi^j_0(L+r_j)(1-i\omega)+2i\omega\overline{\phi^j_0(L+r)}+\\
&+2i\omega r_j\overline{\phi'^j_0(L+r_j)}+\phi^j_{1}(L+r_j) + \Omega^j_{1}(L+r_j)\bigg]
\end{aligned}
\label{eq:integ}
\end{equation}
where $\omega=\lambda'(L)$. The above system can be used to approximate stress intensity factors for any crack pair that extends the initial parallel straight cracks by $\lambda(x)$.  
Finally, to model crack \emph{growth} and determine the actual path $\lambda$ that the freely growing crack will follow, we must select $\lambda$ such that the opening criterion
\begin{equation}\label{eq:stop}
K_{II}^1(L)=K_{II}^2(L)=0 \ \ \text{for } 0<L<L_{stop}
\end{equation}
is always satisfied as the cracks grow to some total extension length $L_{stop}$. 
By solving the system of integral equations (Eq.\ref{eq:fandw}-\ref{eq:stop}) the crack path along with the stress field during growth can be calculated. Based on the results of Cotterell and Rice and Sumi et al. \cite{CotterellRice1980,sumi1983}, we will assume the path is of the general form
\begin{equation}
\lambda(x)=\alpha x+\beta x^{3/2}+\gamma x^2+O(x^{5/2})
\end{equation}
with $\alpha, \beta$, and $\gamma$ constants.  Truncating beyond $x^2$, it is our goal, hence, to solve the above system for these three constants.  While solving the system is still non-trivial, next we propose three solution methods that reduce the integral equations to a system of more tractable algebraic equations.
\section{Method I}
\indent By assuming $\frac{\lambda(L)}{L} \ll 1$, we replace the mirror tractions ($T_n^{(mt)}$ and $T_s^{(mt)}$) by a Taylor series in $\sqrt{x}$ with unknown constants.:
\begin{equation}
T_n^{(mt)}(x_j)-i  T_s^{(mt)}(x_j) = \sum_{n=0}^{\infty}(P_{n/2}-i  Q_{n/2})(\frac{x_j}{L})^{\frac{n}{2}}.
\label{eq:mirrortraction}
\end{equation}
The (truncated) Williams expansion for the stress near the initial cracks is given by
\begin{equation}
\begin{aligned}
\sigma_{xx}(x_j,0)&=\frac{k_I}{\sqrt{2 \pi x_j}}+T+b_I\sqrt{\frac{x_j}{2\pi}}\\
\sigma_{yy}(x_j,0)&=\frac{k_I}{\sqrt{2 \pi x_j}}+b_I\sqrt{\frac{x_j}{2\pi}}\\
\sigma_{xy}(x_j,0)&=\frac{k_{II}}{\sqrt{2 \pi x_j}}+b_{II}\sqrt{\frac{x_j}{2\pi}}\\
\end{aligned}
\label{eq:asymStress}
\end{equation}
where $k_I, k_{II}, b_I, b_{II},$ and $T$ are known Williams expansion coefficients for the initial unextended cracks \cite{isida1970,hori1985}. Mindful that the two close-field tractions  in Eq.\ref{eq:traction1}  are symmetric ($T_s^{1(cf)}=T_s^{2(cf)}\equiv T_s^{(cf)}$ and $T_n^{1(cf)}=T_n^{2(cf)}\equiv T_n^{(cf)}$) we can resolve the above stresses along the extension path to obtain by the assumption of the small deflection (see Eq.17-25 in \cite{sumi1983} or Eq.39-49 in \cite{CotterellRice1980} for more detail).
%\begin{widetext}
\begin{equation}
\hspace{-1cm}\begin{aligned}
&T_n^{(cf)}(x_j) = (k_I-\frac{3}{2}\alpha k_{II})\frac{1}{\sqrt{2\pi x_j}}-\frac{5\beta k_{II}}{2\sqrt{2\pi}}+\\
&\hspace{1cm} +(b_I-\frac{7}{2}\gamma k_{II}-\frac{5}{2}\alpha b_{II})\sqrt{\frac{x_j}{2\pi}} \\
&T_s^{(cf)}(x_j) = (k_{II}+\frac{\alpha}{2}k_I)\frac{1}{\sqrt{2\pi x_j}}+(-\alpha T+\frac{\beta k_I}{2\sqrt{2\pi}})+\\
&\hspace{1cm} +(b_{II}-3\sqrt{\frac{\pi}{2}}\beta T+\frac{\gamma k_I}{2}-\frac{\alpha b_I}{2})\sqrt{\frac{x_j}{2\pi}}\\
\label{eq:Tms}
\end{aligned}
\end{equation}
%\end{widetext}
It bears mentioning that how many terms one keeps in  the Williams expansion of Eq. \ref{eq:asymStress} places an inherent limit on how large $L_{stop}$ can be.  We can only extend the crack as far as the close-field solution is an accurate representation of the stress field adjacent to the crack tips in Problem A.  In Appendix \ref{app:kIext}, we justify our selection of keeping terms up to $x_{j}^{1/2}$ in Eq. \ref{eq:asymStress}.
By using Eq. \ref{eq:mirrortraction} and Eq. \ref{eq:Tms} to define the tractions accordingly in Eq.\ref{eq:fandw}, the functions $\phi$ and $\Omega$ can be derived in terms of the unknown coefficients (Appendix \ref{app:B}, Eq.\ref{equ:field}).
%where $\phi^*_0$ and $\phi^*_1$ are presented in the Appendix \ref{app:B}.
Putting the solution into Eq.\ref{eq:integ} gives us \\
%\begin{widetext}
%\begin{equation}
%\hspace{-1.3cm}\begin{aligned}
%K_{II}&=\left(\frac{\alpha k_I}{2}+k_{II}\right)\\
%&+\bigg(2 \alpha \sum_{n=0}^{N} \frac{\frac{n}{2}! P_{\frac{n}{2}}}{2 \sqrt{2} \left(\frac{n}{2}+\frac{1}{2}\right)!}+\frac{3 \beta k_I}{4}+\frac{3 \alpha  \beta  k_{II}}{2 \pi }-\frac{9 \alpha  \beta  k_{II}}{8}-2 \sqrt{\frac{2}{\pi }} \alpha  T+
%2 \sum _{n=0}^{N} \frac{ \frac{n}{2}! Q_{\frac{n}{2}}}{\sqrt{2} \left(\frac{n}{2}+\frac{1}{2}\right)!}\bigg)\sqrt{L}\\
%&+\bigg(-\frac{\alpha  b_I}{4}+\frac{b_{II}}{2}+\gamma  k_I-\frac{9 \alpha  \gamma  k_{II}}{8}-\frac{5 \beta ^2 k_{II}}{4 \pi }-\frac{3}{2} \sqrt{\frac{\pi }{2}} \beta  T+3 \beta  \sum _{n=0}^{N} \frac{ \frac{n}{2}! P_{\frac{n}{2}}}{2 \sqrt{2} \left(\frac{n}{2}+\frac{1}{2}\right)!}\bigg)L\\
%&+O(L^{3/2})\\
%\end{aligned}\label{KIIM1}
%\end{equation}
%\end{widetext}
\begin{equation}
\begin{aligned}
K_{II}&=\left(\frac{\alpha k_I}{2}+k_{II}\right)\\
&+\bigg(2 \alpha \sum_{n=0}^{N} \frac{\frac{n}{2}! P_{\frac{n}{2}}}{2 \sqrt{2} \left(\frac{n}{2}+\frac{1}{2}\right)!}+\frac{3 \beta k_I}{4}+\frac{3 \alpha  \beta  k_{II}}{2 \pi }-\frac{9 \alpha  \beta  k_{II}}{8}-\\
&-2 \sqrt{\frac{2}{\pi }} \alpha  T+
2 \sum _{n=0}^{N} \frac{ \frac{n}{2}! Q_{\frac{n}{2}}}{\sqrt{2} \left(\frac{n}{2}+\frac{1}{2}\right)!}\bigg)\sqrt{L}\\
&+\bigg(-\frac{\alpha  b_I}{4}+\frac{b_{II}}{2}+\gamma  k_I-\frac{9 \alpha  \gamma  k_{II}}{8}-\frac{5 \beta ^2 k_{II}}{4 \pi }-\\
&-\frac{3}{2} \sqrt{\frac{\pi }{2}} \beta  T+3 \beta  \sum _{n=0}^{N} \frac{ \frac{n}{2}! P_{\frac{n}{2}}}{2 \sqrt{2} \left(\frac{n}{2}+\frac{1}{2}\right)!}\bigg)L+O(L^{3/2})\\
\end{aligned}\label{KIIM1}
\end{equation}
Since Eq.\ref{eq:stop} requires that the above vanish  for all $0<L<L_{stop}$ it follows that the coefficients of each power of $L$ in the above must vanish. That is, each term in parenthesis must vanish, which adds three more equations to Eq.\ref{eq:NEqu}, for a total of $2N+3$ nonlinear equations.  Solving this system,  we obtain the Taylor series constants along with the coefficients $\alpha$, $\beta$, and $\gamma$, which give the solution for the crack path.  This concludes what we refer to as `Method 1' for computing $\lambda(x)$.
To grow cracks a very long distance, a key limitation is the accuracy of the selected close-field solution.  Because we prefer the convenience of our truncation in Eq \ref{eq:asymStress}, future work will explore an iterative approach to overcome this issue by growing the crack in a piecewise sequence, using the stress at the end of a growth increment to recalculate new close-field coefficients for the next step of growth. 
\section{Method II} 
In the case of $L_{stop}\ll r_0, a $, the $P$ and $Q$ coefficients in Eq.\ref{KIIM1} can be neglected 
 resulting in a closed form solution for the constants in $\lambda(x)$: 
\begin{equation}
\begin{aligned}
&\alpha =-\frac{2k_{II}}{k_I}\\
&\beta=-\frac{16 \sqrt{2 \pi } k_{II} T}{3 \left(\pi  k_I^2+(3 \pi -4) k_{II}^2\right)}\\
&\gamma=\frac{8}{8k_I-9\alpha k_{II}}\bigg(\frac{\beta^2 k_{II}}{4\pi}+\frac{3}{2}\sqrt{\frac{\pi}{2}}\beta T+\frac{\alpha b_I}{4}-\frac{b_{II}}{2}\bigg)\\
\end{aligned}
\label{eq:alphas}
\end{equation}
The above formulae constitute `Method II' for approximating EP crack paths; it is merely a simplification imposed on  the system from Method I. The above result, when linearized in $\alpha$ and $\beta$, is compatible with the first-increment path solution of Sumi et al. \cite{sumi1983}.   Method II effectively ignores the influence of crack 2's extension on the path taken by crack 1, which is justifiable for small extensions. 
\section{Validation of Method I and Method II}
\begin{figure}[t]
\centering
\includegraphics[width=0.4\textwidth]{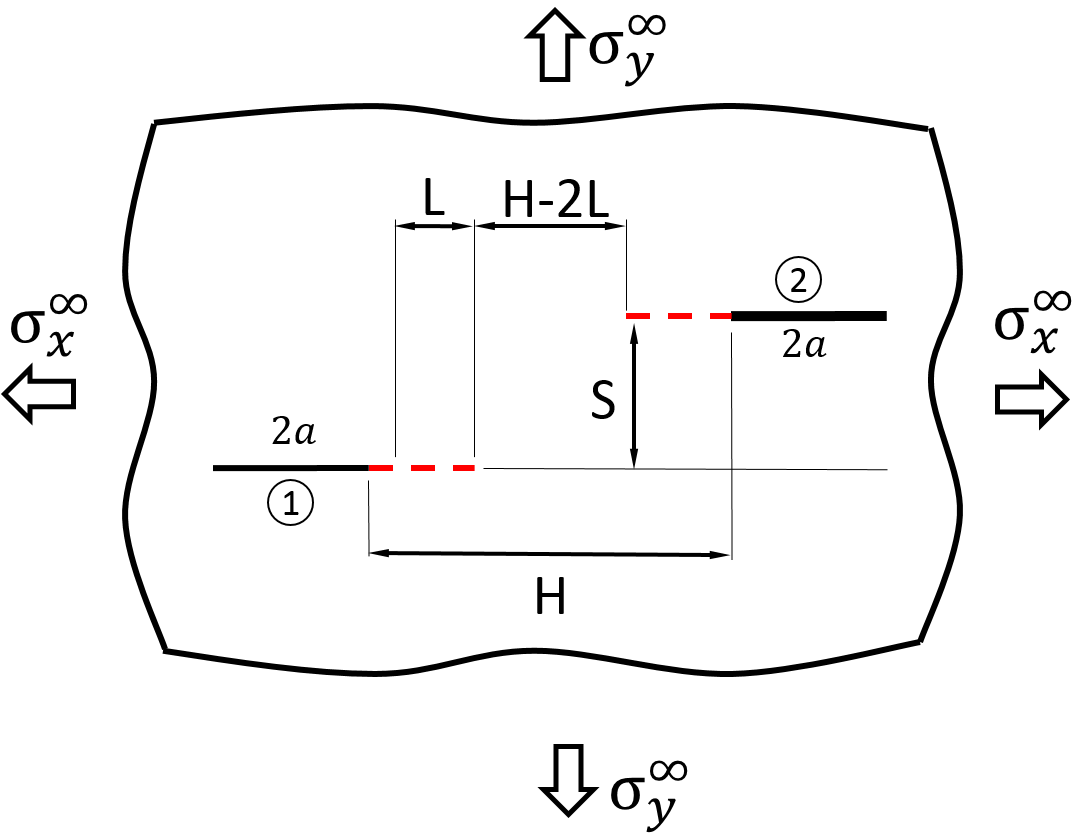}
\caption{Verification test for Method I and Method II.  We use the methods to approximate the value of $K_{II}$ after an imposed extension of length $L$. The solid line shows the main crack and dashed lines are the extensions. 
}
\label{fig:verification}
\end{figure}
The accuracy of the two above methods depends on how well they represent the stress fields near a pair of extending crack tips for any  path $\lambda$.  In order to validate the simplifications used in deducing these methods,  a numerical experiment has been performed to test each method's underlying stress prediction in a case with an exact solution against which to compare.  As it is shown in Fig.\ref{fig:verification} we suppose cracks number 1 and number 2 begin with the same length of $2a$ and are positioned such that their tips are separated horizontally by $H$ and vertically by $S$.  We now apply a straight line extension of length $L$ to both cracks, i.e. $\lambda=0$, and determine $K_{II}$ based on the fields $\phi_0$, $\Omega_0$, $\phi_1$, and $\Omega_1$ as they are given in Method I and as they are further approximated in Method II. 
We choose $\sigma_x^{\infty}=0$ for these tests.  Our estimations are then compared to the known exact result for $K_{II}$ for a pair of straight cracks of length of $2a+L$. The results are presented in Tab.\ref{tab:veri}.
\begin{table}
\small
\caption{Verifying the approximation of $K_{II}$ (in units of $\sigma_{\infty}\sqrt{a}$) after extension using Method I and Method II and comparing to the exact solution. All lengths in units of $a$.}
\begin{center}
\begin{tabular}{|| c | c | c || c | c || c | c ||}
\hline
 \multirow{2}{*}{$H$, $L$}  & \multirow{2}{*}{$S$}  & Exact   & \multicolumn{2}{c||}{Method I} & \multicolumn{2}{c||}{Method II} \\ \cline{3-7}
	  &	   & $K_{II}$& $K_{II}$ & Err(\%) & $K_{II}$ & Err(\%) \\ \hline
\multirow{3}{*}{10, 1}  & 6  & -0.0010 & -0.0010  & $<$1       & 0.000   & $>$100 \\
  & 10 & -0.0117 & -0.0116  & $<$1    & -0.0050    & 70 \\
  & 14 & -0.0110 & -0.0106  & 3       & -0.0055    & 50 \\ \hline
\multirow{3}{*}{5, 0.2} & 6  & -0.0140 & -0.0135 & 3 & -0.0116 & 17 \\
  & 10 & -0.0130 & -0.0124 & 5 & -0.0112 & 13 \\
  & 14 & -0.0081 & -0.0080 & 1 & -0.0070 & 13 \\ \hline
\multirow{3}{*}{5, 0.1}  & 6  & -0.0120  & -0.0113 & 6 & -0.0108 & 10 \\
  & 10 & -0.0117  & -0.0117 & $<$1 & -0.0108 & 8 \\
  & 14 & -0.0070 & -0.0070 & $<$1 & -0.0067 & 4 \\ \hline
\end{tabular}
 \end{center}\label{tab:veri}
\end{table}
\normalsize
The first row of the Table indicates that Method I is able to approximate $K_{II}$ better than Method II, as expected.  It retains accuracy for moderately long extensions --- here, $L$ on the order of $a$ --- while Method II loses almost all accuracy in this range.  For small crack extensions, the second and third rows of the Table, the difference in accuracy of the two methods is much smaller. We observe sufficient agreement between both methods and the exact result as the crack extension length decreases, and the accuracy of both methods tends to increase when the initial cracks are farther apart.

\begin{figure}[!ht]
\centering
\includegraphics[width=0.45\textwidth]{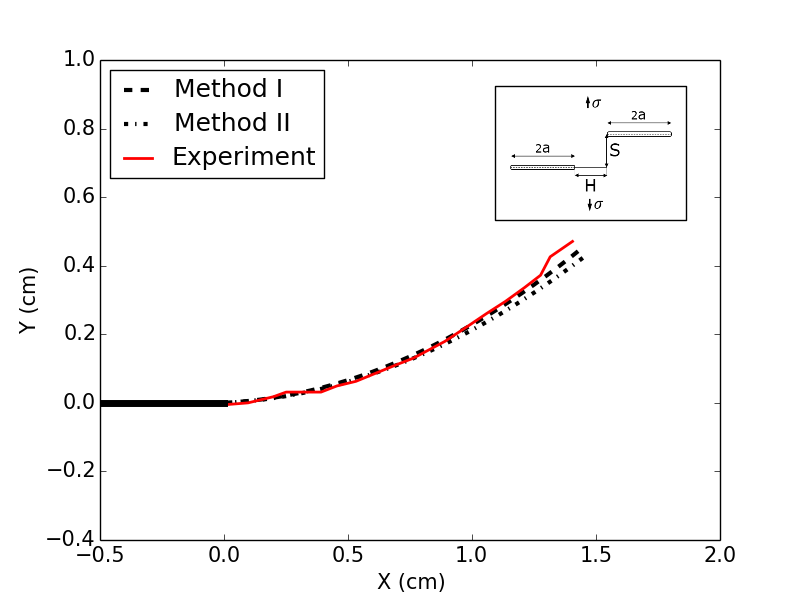}
\caption{Comparison of the predicted crack path shape against the average experimentally observed path \cite{cortet2008} for a pair of cracks of length $2a=1$cm positioned with $H=1$cm and  $S=$1.6cm .}
\label{fig:excom}
\end{figure}

\rcom{Cortet et al \cite{cortet2008} conducted an experimental study of  the growth of a periodic array of cracks in a thin paper material.  The crack arrangement is such that neighboring crack pairs sufficiently replicate the EP-crack geometry.  Their data provides a statistically well-averaged crack path shape that can be used as a preliminary experimental check on our model.    Fig. \ref{fig:excom} compares our predicted path shape for $S=1.6$ cm $ \approx 3.2a$ and $H=1$ cm $\approx 2a$ to the experimentally observed average path shape of this configuration.  Strong agreement is observed. Other configurations with the same $H$ but smaller $S$ values were studied in that work, but paths presented were preferentially averaged over different subcategories of behavior, to which we cannot apply our model.  Based on a probabilistic graph presented in that work, when $S=0.6$cm$\approx 1.2a$ and $H=1$cm$\approx 2a$ there is a $50\%$ chance for both repulsion and attraction between the cracks. One might interpret this as an indication that the kink angle is on-average zero for this configuration.   Though the crack positioning may be approaching the closeness limits of our models, we have found that at the same $H$, Methods I and II both predict the kink angle to vanish at $S\approx0.9$cm, which is not perfect but in the same range as the experimental result.  These are a preliminary experimental checks; a more in-depth experimental study using different materials and  separation protocols, and thicker out-of-plane dimensions will be needed to further validate our models.}
%In the case of crack path, Fig. \ref{fig:excom} shows the comparison of our prediction for $S=1.6 (cm) \approx 3.2a$ and $H=1 (cm)\approx 2a$ \footnote{Except for $d=1.6 (cm)$ no other global average is presented based on the separation distance; i.e. all the cases are averaged in two separated cases of repulsion and attraction. Meanwhile, $d=1.6 (cm)$ is the mostly attraction based their probabilistic results}. More experiments with a specific assumption we have made are needed obviously for the validation of the mathematical model; anyhow, the prilimanary comparison shows a very good correspondence with the experiments.
\section{Method III:  Asymptotic matching}\label{matching}
The methods above make reference to a close-field solution based on a known analytical solution for Problem $A$. This analytical solution is itself not trivial to obtain and lengthy to write  \cite{isida1970, yokobori1971, hori1985}.  We are left to wonder if there is a simpler way to approximate the close-field solution's Williams coefficients, so that Method II could be employed more quickly.  If such a way existed, it would also serve potential benefits as we attempt to model other multi-crack systems in the future, for which analytical solutions of the initial problem may not be tenable.

Consider a single Griffith's crack under bi-axial mode-one loading.  In the general case, based on Westergaard's solution \cite{westergaard1939}, the $k_I$, $b_I$, and $T$ values are non-zero while $k_{II}$ and $b_{II}$ have zero values. If we now introduce a second crack in the loaded system, the deviations in the Williams coefficients from the single crack solution go as $r_0^{-2}$, as we show in Appendix \ref{app:C}.  Since $k_I$, $b_I$, and $T$ were initially finite of order one, the deviation induced by the second crack does not affect their leading order behavior (in $r_0^{-1}$), however since $k_{II}$ and $b_{II}$ were initially zero, their behavior is governed entirely by the presence of the second crack.  Then, by applying Eq. \ref{eq:asymStress}, we find that the leading order behavior of $\alpha$, $\beta$, and $\gamma$  are unaffected by the perturbations to $k_I$, $b_I$, and $T$ induced by the second crack, but depend directly on the perturbations to $k_{II}$, and $b_{II}$.  Likewise, by calculating the perturbed $k_{II}$ and $b_{II}$ values and using them along with the single-crack values of $k_I$, $b_I$, and $T$, we can then approximate the EP crack path under Eq \ref{eq:asymStress}.

\begin{figure}[t]
\centering
\includegraphics[width=0.48\textwidth]{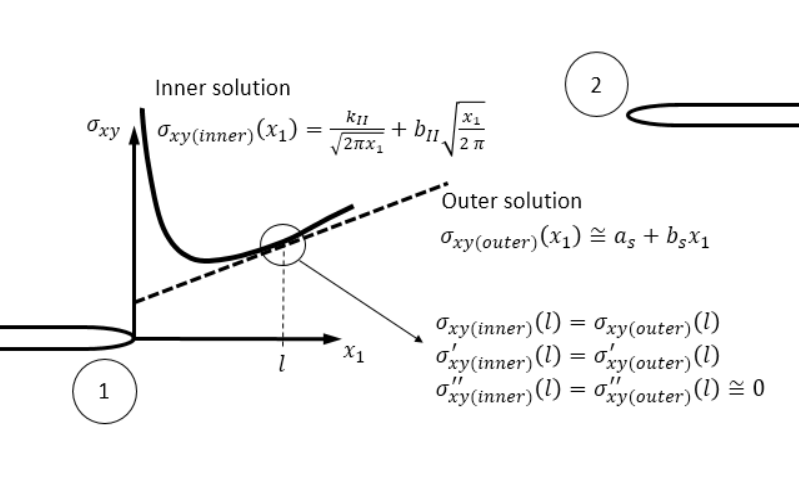}
\caption{Matching the inner solution based on the Williams expansion stress field, to the far-field solution given by the superposition of single-crack fields.
}
\label{fig:sumcrackschem}
\end{figure}
\begin{figure}
\centering
\includegraphics[width=0.45\textwidth]{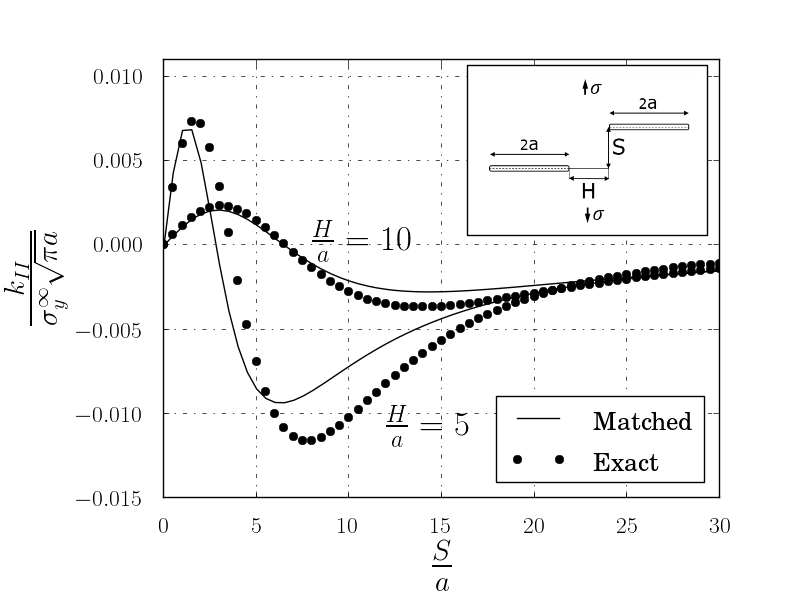}
\caption{The value of $k_{II}$  for EP cracks obtained by matching the asymptotic fields compared to the exact solution.}
\label{fig:kIIcomp}
\end{figure}

The following is a novel approach to approximate $k_{II}$ and $b_{II}$, by fitting them from an ``outer solution'', i.e. a stress field for the two-crack problem that is accurate not close to either crack.  There are potentially many ways to obtains an outer solution, but herein we discuss one such way.  A more complete analysis of the method --- including closed-form error estimates for the outer solution, and an asymptotically valid $k_{II}$ approximation for EP cracks (Eq. \ref{asympk})--- are included in Appendix \ref{app:C}.

The exact solution to the biharmonic function $\phi$ for our Problem $A$ (ignoring the constant far-field stress) can be expressed as the sum of the solutions for two isolated Griffith cracks along with an extra part $\phi_{res}(z)$ which comes from the interaction of each crack on the other one. As it is expected, if one neglects $\phi_{res}$ the resulting form is not able to capture the stress intensity factors of the cracks but does a sufficient job representing the stress field when not close to either crack (see Appendix \ref{app:C}).  Therefore, in a region that excludes the vicinity of either crack, an ``outer solution'' for the stress field can be assumed to be a superposition of two separated single cracks under tension.
 Meanwhile, the stress near the crack tip has to always follow a Williams expansion. Consequently, requiring that an asymptotic inner solution ---  taken to be a truncated Wiliams expansion --- match the outer solution in some overlap window could yield a fast method to produce the needed inputs in Eq. \ref{eq:alphas} to obtain $\lambda$.  \footnote{Unlike standard asymptotic matching techniques which involve a small parameter that induces a rapidly varying inner solution expressible in stretched coordinates, the inner solution here is rapidly varying because it is a true singular function and does not require stretched coordinates \cite{holmes2013}.}
\begin{figure*}[t]
\centering
\subfigure[][$S=5a$ and $H=10a$]{\includegraphics[width=0.41\textwidth]{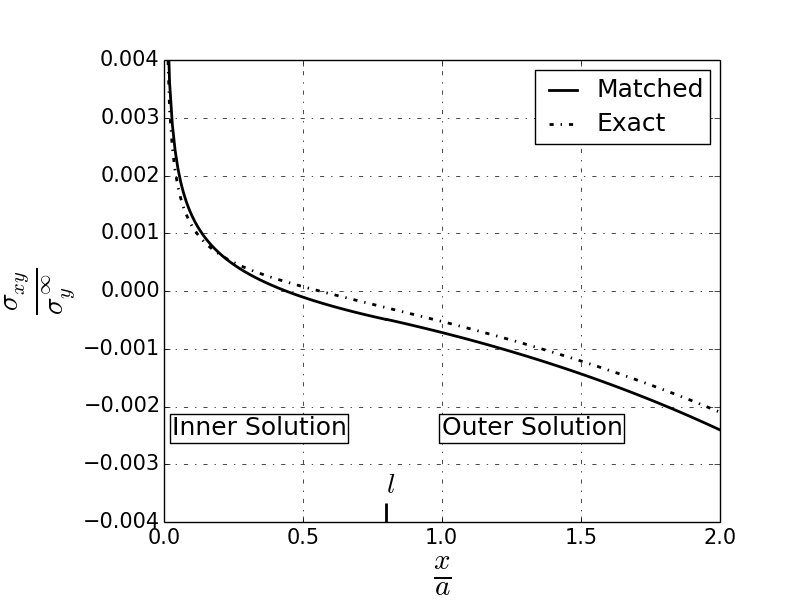}}
\subfigure[][$S=10a$ and $H=10a$]{\includegraphics[width=0.41\textwidth]{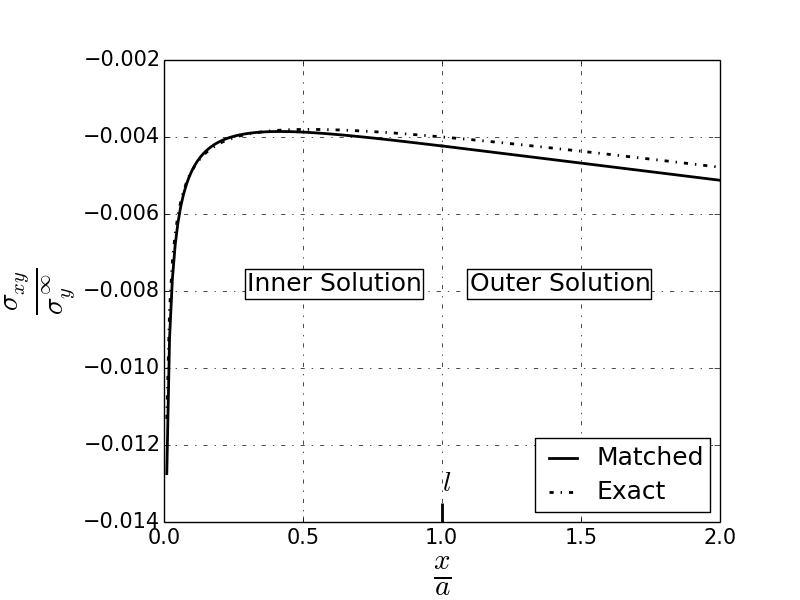}} \\
\subfigure[][$S=5a$ and $H=20a$]{\includegraphics[width=0.41\textwidth]{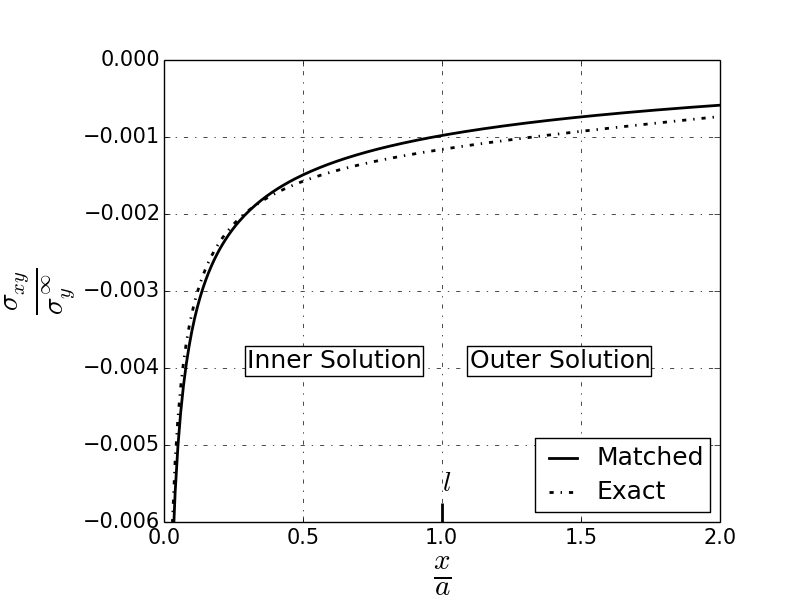}}
\subfigure[][$S=15a$ and $H=5a$]{\includegraphics[width=0.41\textwidth]{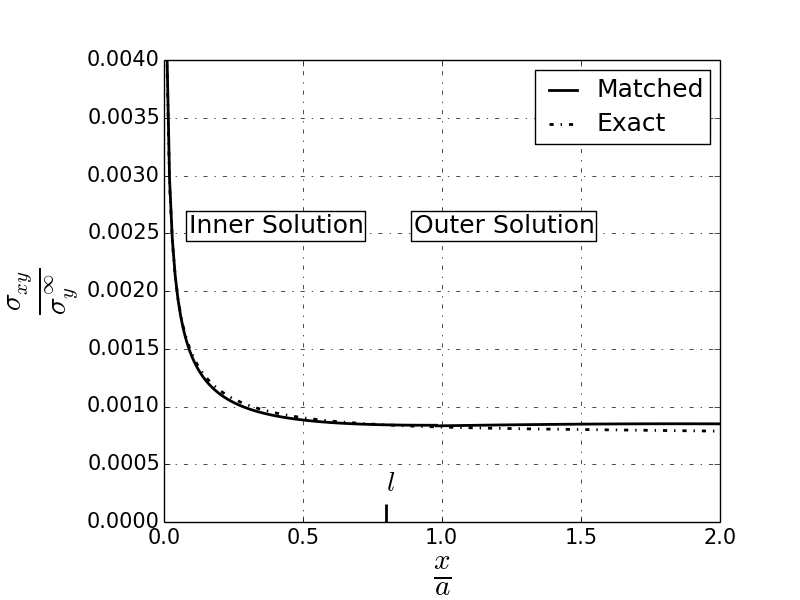}} 
\caption{Comparison of the asymptotically matched and exact solutions for $\sigma_{xy}(x,y=0)$, for $x=0$ at the lower-left crack tip, for (a) $S=5a$ and $H=10a$ and (b) $S=10a$ and $H=10a$ (c) $S=5a$ and $H=20a$ (d) $S=15a$ and $H=5a$. The dashed vertical line shows the matching point $l$ as presented in Eq. \ref{eq:optequ}}\label{fig:MatchedH10}
\end{figure*} 
Let $\mathbf{\sigma}^G(z_1)+\sigma_x^{\infty}\mathbf{e}_x\otimes\mathbf{e}_x+\sigma_y^{\infty}\mathbf{e}_y\otimes\mathbf{e}_y$ represent the stress field for a single Griffiths crack of length $2a$ under tensile loading $\sigma_{y}^{\infty}$ and lateral loading $\sigma_{x}^{\infty}$, where $z_1$ originates at the right-side crack tip.  The Griffith's field $\mathbf{\sigma}^G$ is the variation from the far-field stress. The $xy$ shear stress has the exact solution
\begin{equation}
\begin{aligned}
\sigma_{xy}(x_1)&=\sigma^G_{xy}(z_1) = -2 Im(z_1) Re(\phi^{G\hspace{0.05cm}\prime}(z_1))\\
\phi^G(z_1)&=\frac{\sigma_y^{\infty}(z_1+a)}{2\sqrt{(z_1+a)+a}\sqrt{(z_1+a)-a}}-\frac{\sigma_y^{\infty}}{2}.
\end{aligned}
\label{eq:slitstress}
\end{equation}
To construct an outer solution for twin cracks, we suppose a superposition of the two Griffiths fields, as if each crack were on its own.  Hence,
\begin{equation}\label{sumcrackstress}
\begin{aligned}
&\mathbf{\sigma}_{outer}(z_1)=\mathbf{\sigma}^G(z_1)+\mathbf{\sigma}^G(z_2)+\sigma_x^{\infty}\mathbf{e}_x\otimes\mathbf{e}_x+\sigma_y^{\infty}\mathbf{e}_y\otimes\mathbf{e}_y
\\
&\ z_2=z_0-z_1, \ \ z_0=r_0 e^{i \theta_0}
\end{aligned}
\end{equation}
To perform the matching, we choose to enforce agreement between the inner and outer solutions at some point $(l,0)$ on the line $y_1=0$.  We choose this line because the function $\sigma_{xy}(x_1\to 0, y_1=0)$ is decoupled entirely from the mode-I crack-tip coefficients.  Thus the behavior of $\sigma_{xy}$ on this line permits us to extract the needed $k_{II}$ and $b_{II}$ values directly without having to filter out dependences on the much-larger mode I coefficients, a process that would introduce new sources of error. 
\begin{figure*}[t!]%{l}{0.45\textwidth}
\centering
\subfigure[][$H/a=5$]{
\includegraphics[width=0.3\textwidth]{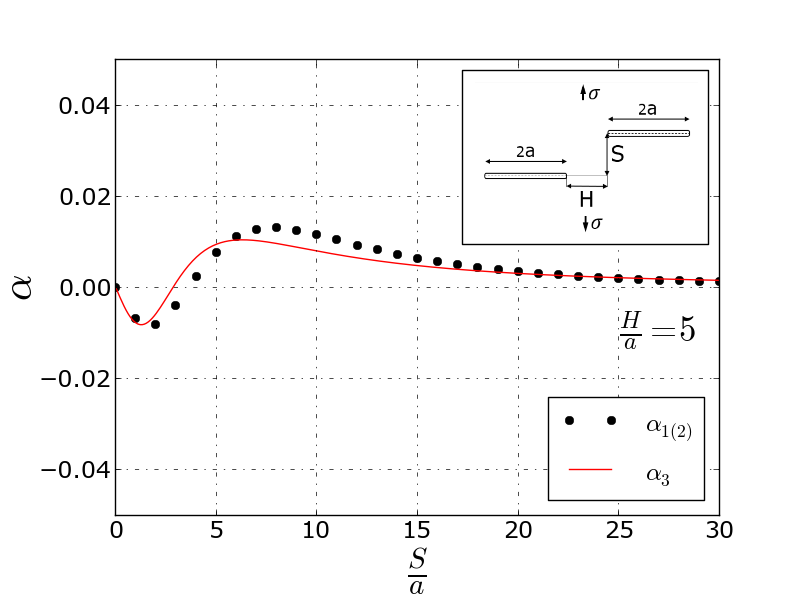}
\includegraphics[width=0.3\textwidth]{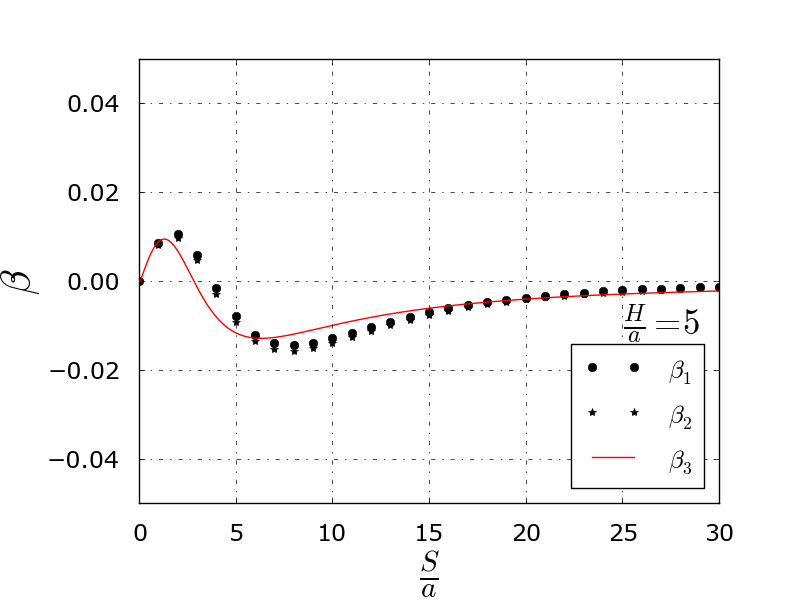}
\includegraphics[width=0.3\textwidth]{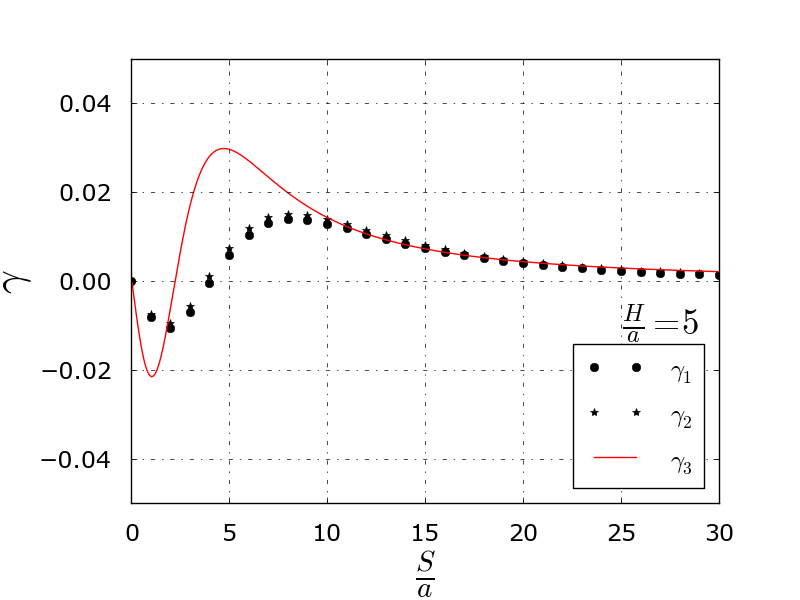}}\\
%%
%\subfigure[][$H/a=10$]{
%\includegraphics[width=0.3\textwidth]{Figures/H10Alpha.png}
%\includegraphics[width=0.3\textwidth]{Figures/H10Beta.png}
%\includegraphics[width=0.3\textwidth]{Figures/H10Gamma.png}}\\
%%
\subfigure[][$H/a=15$]{
\includegraphics[width=0.3\textwidth]{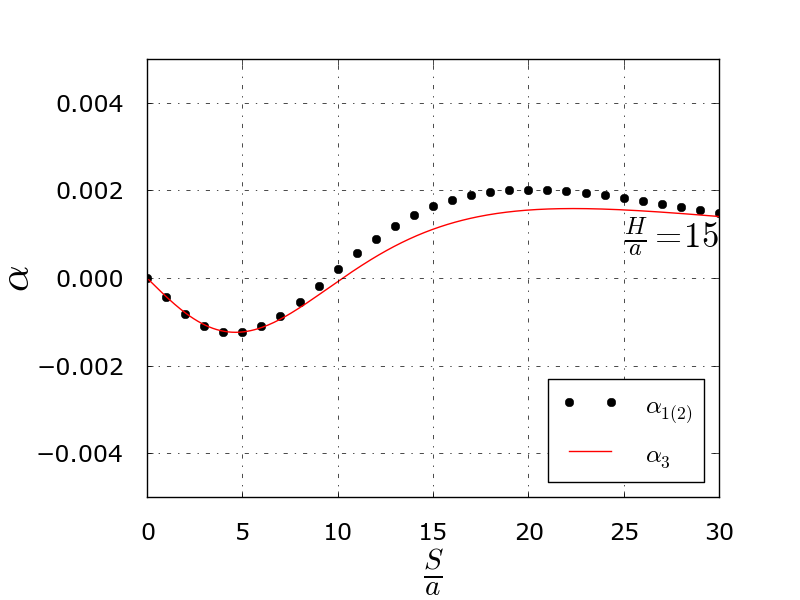}
\includegraphics[width=0.3\textwidth]{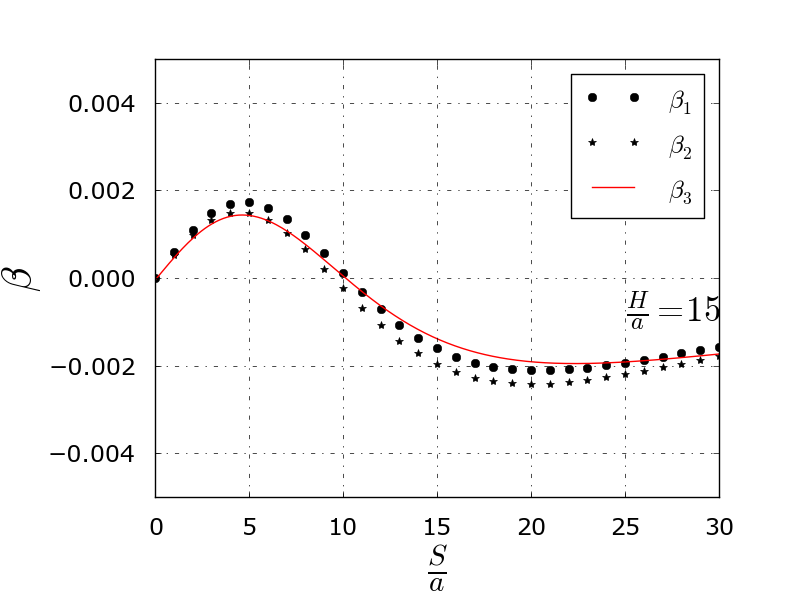}
\includegraphics[width=0.3\textwidth]{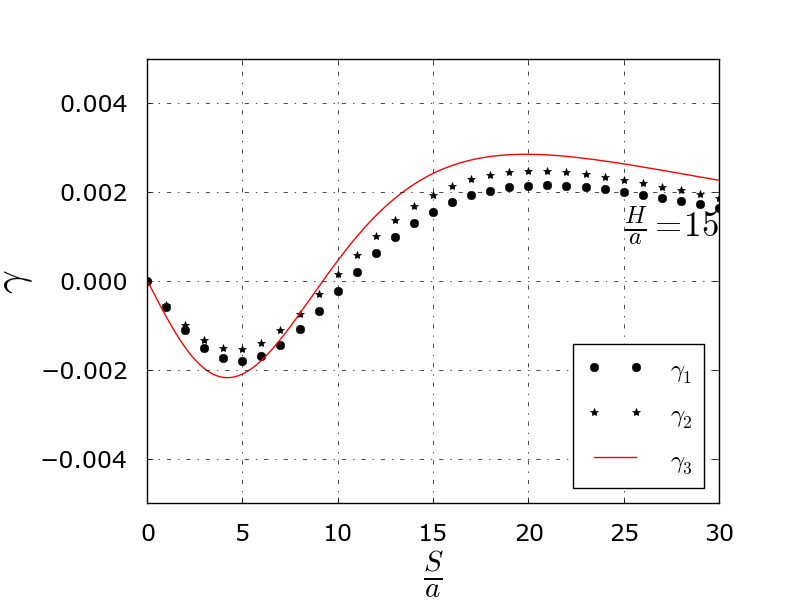}}\\
%%
% \caption{Comparison of the prediction for the constants in $\lambda(x)$ for varying  $S=r_0\sin(\theta_0)$ and constant $H=r_0 \cos(\theta_0)$. (a) $H/a=5$, (b) $H/a=10$, (c) $H/a=15$}
          \label{fig:Hcompare}
          \vspace{-10pt}
%\end{figure*}
%\begin{figure*}
\centering
\subfigure[][$S/a=5$]{
\includegraphics[width=0.3\textwidth]{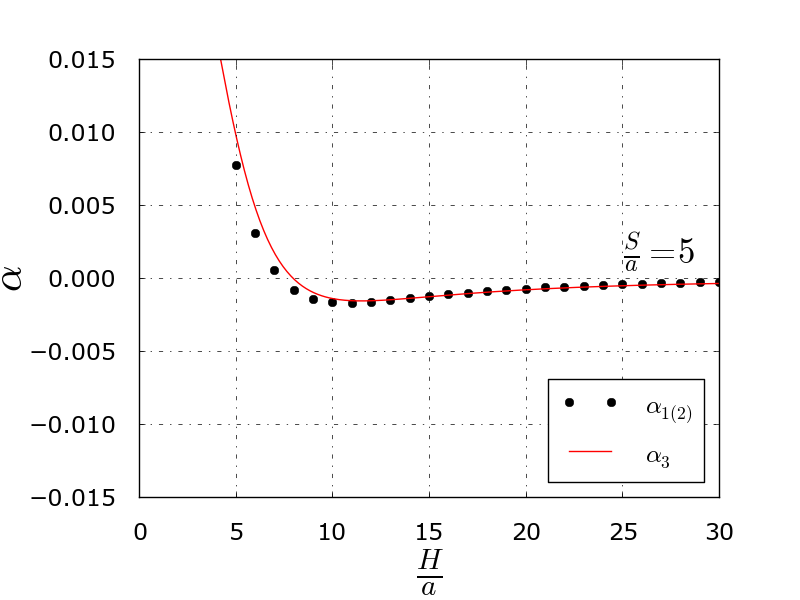}
\includegraphics[width=0.3\textwidth]{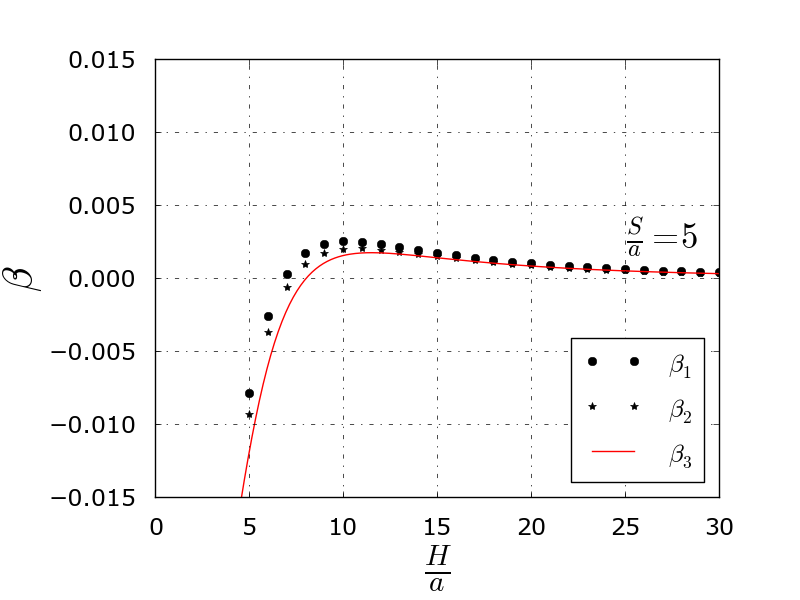}
\includegraphics[width=0.3\textwidth]{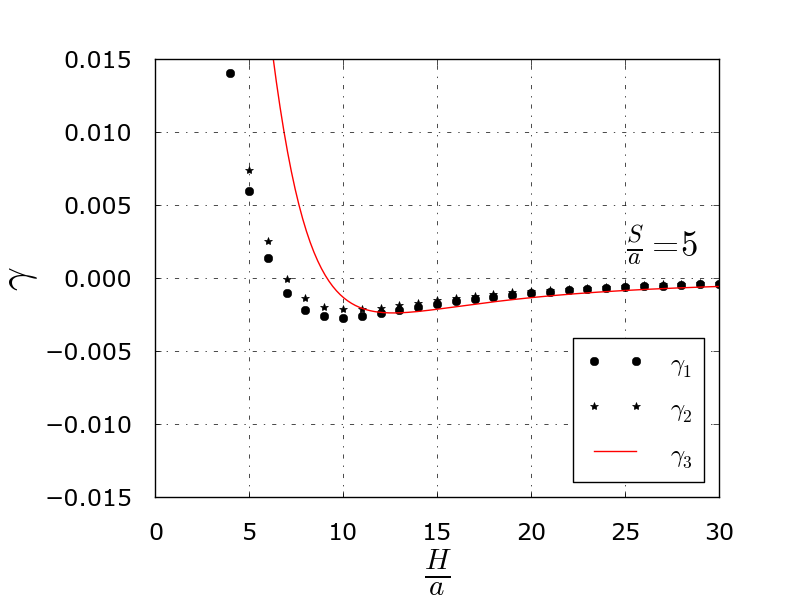}}\\
%%
%\subfigure[][$S/a=10$]{
%\includegraphics[width=0.3\textwidth]{Figures/S10Alpha.png}
%\includegraphics[width=0.3\textwidth]{Figures/S10Beta.png}
%\includegraphics[width=0.3\textwidth]{Figures/S10Gamma.png}}\\
%
\subfigure[][$S/a=15$]{
\includegraphics[width=0.3\textwidth]{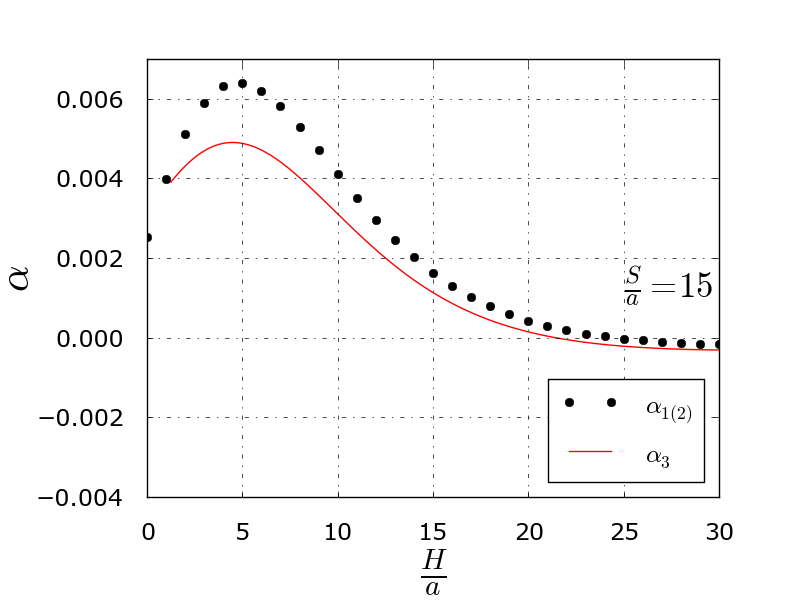}
\includegraphics[width=0.3\textwidth]{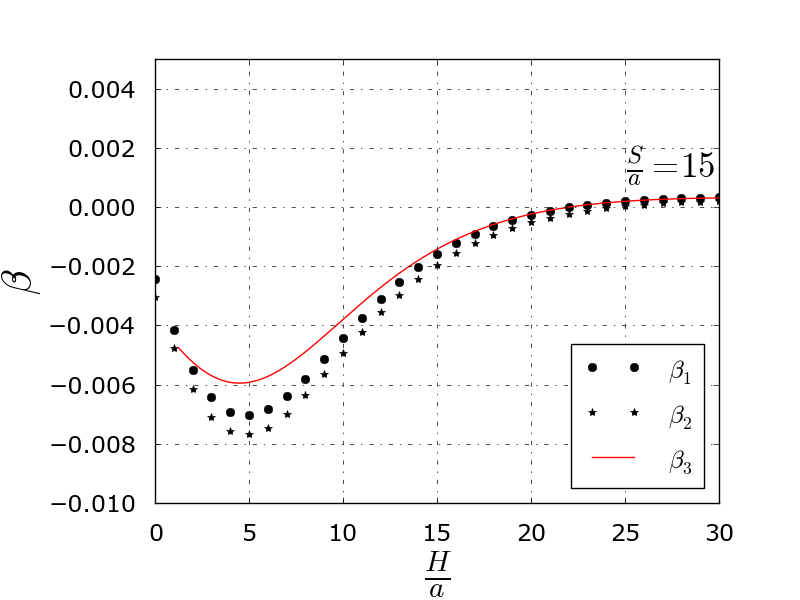}
\includegraphics[width=0.3\textwidth]{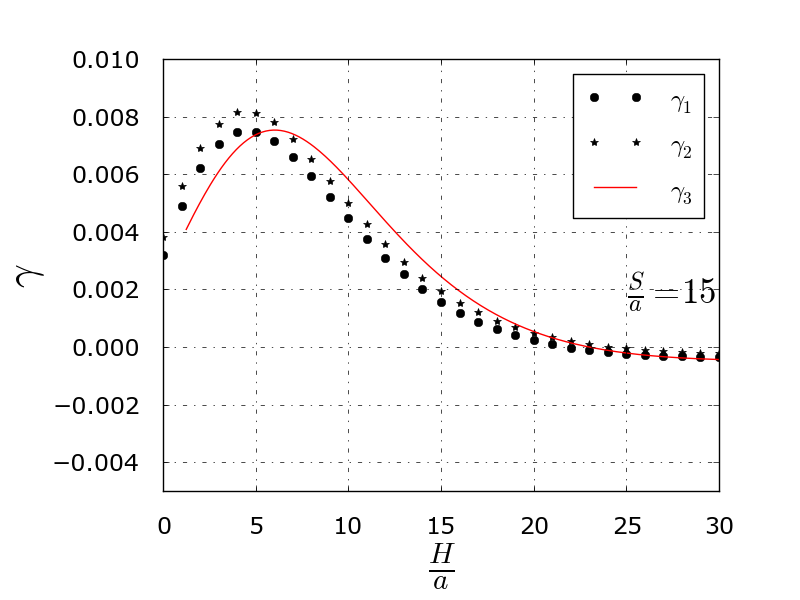}}
%    \caption{Comparison of the prediction for the constants in $\lambda(x)$ for varying $H=r_0 \cos(\theta_0)$ and constant $S=r_0 \sin(\theta_0)$. (a) $S/a=5$, (b) $S/a=10$, (c) $S/a=15$.}
    \caption{Comparison of the prediction for the constants in $\lambda(x)$ for varying $H=r_0 \cos(\theta_0)$ and constant $S=r_0 \sin(\theta_0)$. (a) $H/a=5$, (b) $H/a=15$, (c) $S/a=5$, (d) $S/a=15$.}

          \label{fig:Scompare}
          \vspace{-10pt}
\end{figure*}
Naturally, the field $\sigma^G_{xy}$ does not have a singularity at either crack tip.  However, the outer solution has a nonsingular but finite shear stress at each crack tip, which can be expanded to second-order in a Taylor series per Eq.\ref{eq:optequ}. Focusing on crack 1, an overlap point between the inner and outer solutions, at some $x_1=l$, should have the property that the solutions look similar in a small window about that point.  To identify this point $l$ and simultaneously find the needed constants in the Williams expansion ($k_{II}$ and $b_{II}$)
we desire a matching up to the second-derivative in space between the two solutions at $l$. Hence, by solving the system of equations presented in Eq.\ref{eq:optequ}, the parameters of the inner solution  can be expressed based on the constants in $\sigma_{xy(outer)}(x_1)$. That is,
%%
%\begin{figure}[h]
%\centering
%\includegraphics[width=0.48\textwidth]{Figures/kIIComparison.png}
%\caption{The value of $k_{II}$  for EP cracks obtained by matching the asymptotic fields compared to that of the exact solution.}
%\label{fig:kIIcomp}
%\end{figure}
%
\begin{equation}
\begin{aligned}
&\sigma_{xy(outer)}(x_1)\cong a_{s}+b_{s}x_1+c_s x_1^2; \\ %c_s\ll min(a_{s},b_{s})\cong 0\\
&\sigma_{xy(inner)}(x_1) = \dfrac{k_{II}}{\sqrt{2\pi x_1}}+b_{II}\sqrt{\dfrac{x_1}{2 \pi}}\\
&\begin{cases}
\sigma_{xy(inner)}(l)=\sigma_{xy(outer)}(l)\\ \sigma'_{xy(inner)}(l)=\sigma_{xy(outer)}'(l)\\
\sigma''_{xy(inner)}(l)=\sigma_{xy(outer)}''(l)
\end{cases}
%%\Longrightarrow k_{II}=\frac{1}{3} a_s \sqrt{\frac{2\pi a_s}{3b_s}};
%%\hspace{0.5cm}
%%b_{II}=b_s \sqrt{\frac{6\pi a_s}{b_s}}\\
\end{aligned}
\label{eq:optequ}
\end{equation}

\begin{figure*}[t!!!]
\subfigure[][]{\includegraphics[width=0.33\textwidth]{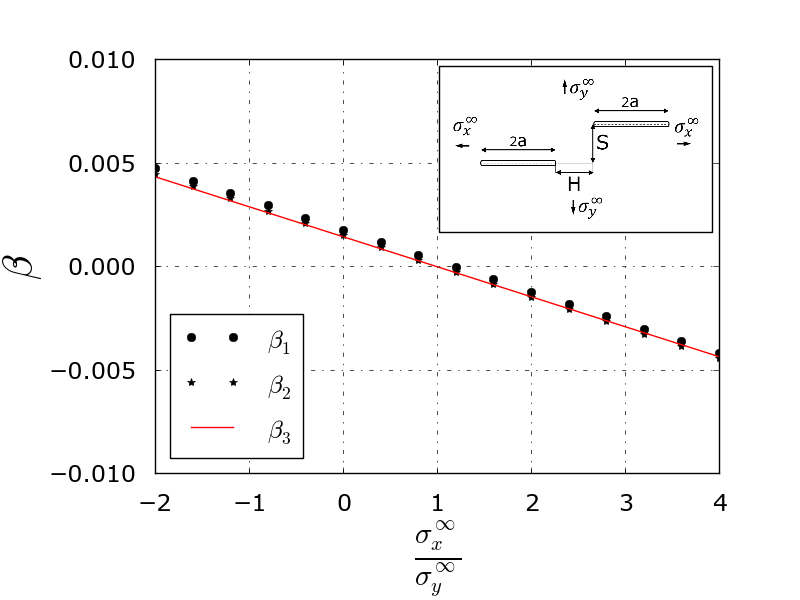}\\
             \includegraphics[width=0.33\textwidth]{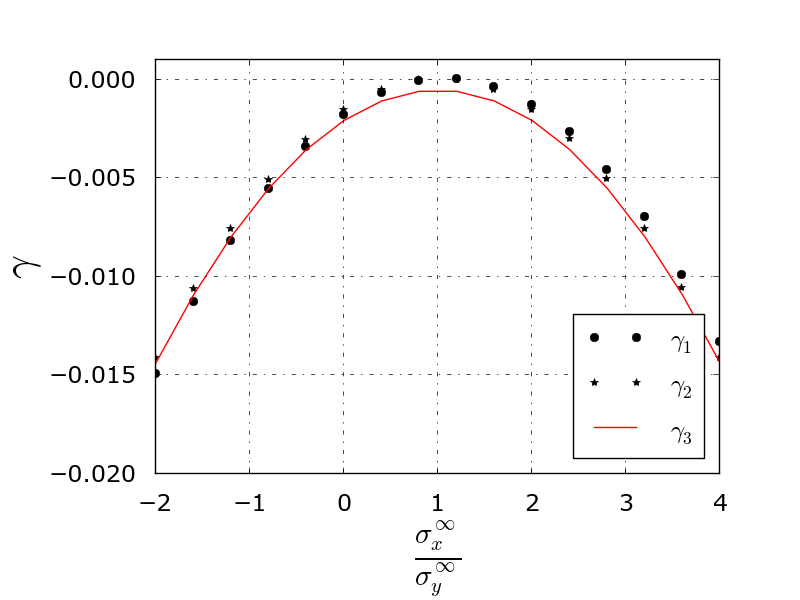}\\
             \includegraphics[width=0.33\textwidth]{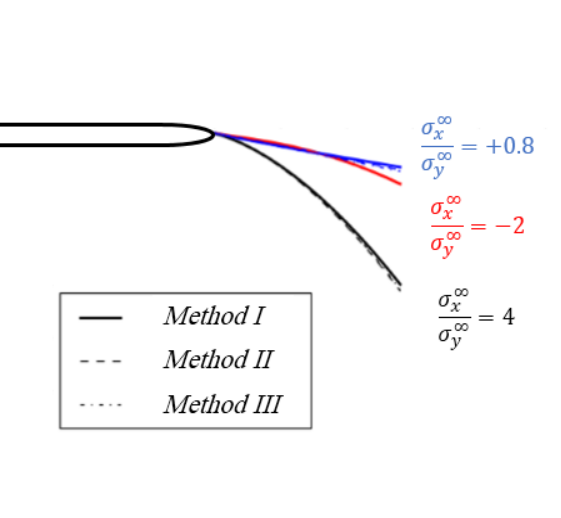}}\\
\subfigure[][]{\includegraphics[width=0.33\textwidth]{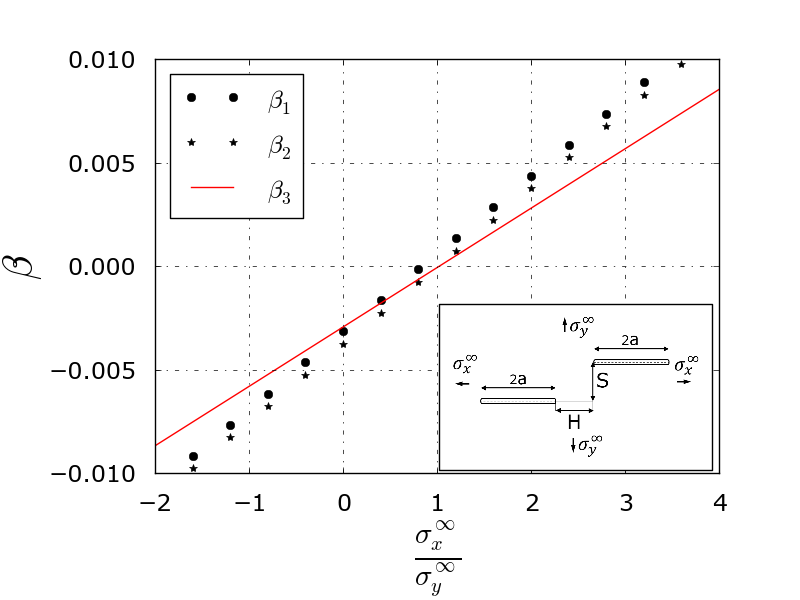}\\
             \includegraphics[width=0.33\textwidth]{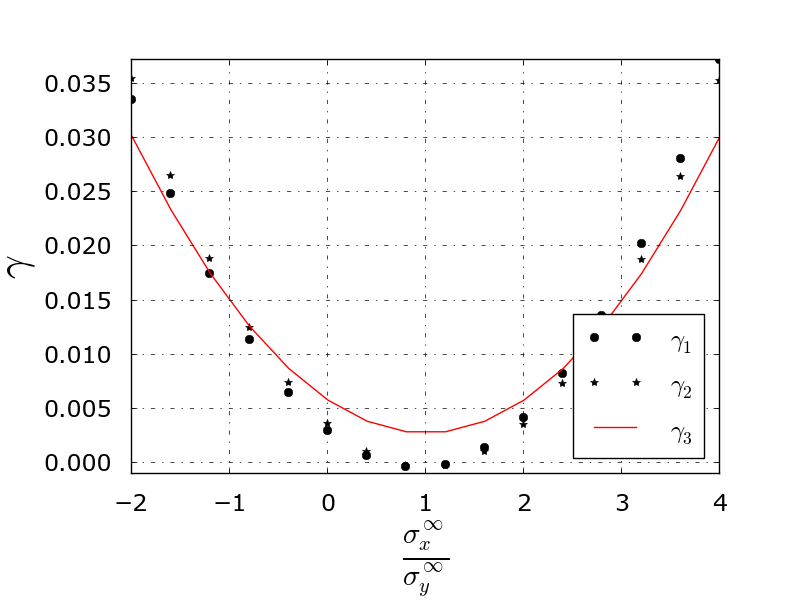}\\
             \includegraphics[width=0.33\textwidth]{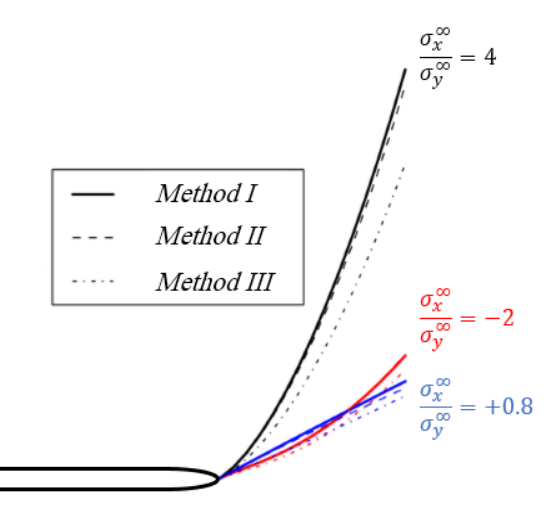}}\\
\caption{Behavior of $\beta$, $\gamma$, and the lower left crack's propagation path for various $\frac{\sigma_x^{\infty}}{\sigma_y^{\infty}}$. Initial crack placements are (a) $H=15a$ and $S=5a$ and (b) $H=10a$ and $S=10a$.}
          \label{fig:TsBetGam}
          \vspace{-10pt}
\end{figure*}
This method  is explained graphically in Fig.\ref{fig:sumcrackschem}.  In the case of EP cracks, for $r_0 \gg a$, we have observed that $c_s\ll min(a_{s},b_{s})$ and thus the constants $l$, $a_s$, and $b_s$ can be solved for directly.  Of particular note, we find the formulas
\begin{equation}
\begin{aligned}
a_s&=-\frac{a^2 \sigma_y^{\infty} Im(z_0) \sqrt{z_0 (z_0-2 a)}}{z_0^2 (z_0-2 a)^2}, \ \ b_s=\frac{3 a^2 \sigma_y^{\infty} Im(z_0) (a - z_0)}{(z_0 (-2 a + z_0))^{5/2}}
\end{aligned}
\label{eq:asbs}
\end{equation}
in which $z_0=r_0e^{i \theta_0}$.
Using Eq.\ref{eq:optequ} and \ref{eq:asbs}, we find the formula
\begin{equation}
\begin{aligned}
k_{II}&=\sqrt{2 \pi } a^2 \sigma_y^{\infty} Im(z_0) Re\left(\frac{ \sqrt{\frac{z_0 (2 a-z0)}{a-z_0}} \sqrt{z_0 (z_0-2 a)}}{9 z_0^2 (z_0-2 a)^2}\right)\\
b_{II}&=-3 \sqrt{2 \pi } a^2 \sigma_y^{\infty} Im(z_0) Re\left(\frac{ (a-z_0) \sqrt{\frac{z_0 (2 a-z_0)}{a-z_0}}}{(z_0 (z_0-2 a))^{5/2}}\right)
\end{aligned}
\label{eq:kIIsum}
\end{equation}
This constitutes `Method III' for EP crack path determination.

In order to verify the precision of the matched solution, Fig.\ref{fig:kIIcomp} shows the value of $k_{II}$ by the matched method, Eq. \ref{eq:kIIsum}, compared to the exact solution for different EP geometries.  Here we use $\sigma_{x}^{\infty}=0$. This figure clearly shows that Eq.\ref{eq:kIIsum} can approximate the exact value of $k_{II}$ and provides a good estimation of the turning point (when the value of the $k_{II}$ changes sign), which depends on the different positions of the cracks.  As another verification, Fig. \ref{fig:MatchedH10} compares the asymptotically matched solution for shear stress, $\sigma_{xy}$, to the exact solution in four different cases. ``Inner solution" and ``Outer solution" regions are matched at the overlap location shown marked with an $l$ on the graphs. 

%
%\begin{figure*}[t]
%\centering
%\subfigure[][$S=5a$ and $H=10a$]{\includegraphics[width=0.35\textwidth]{Figures/MatchedS5-2.png}}
%\subfigure[][$S=10a$ and $H=10a$]{\includegraphics[width=0.35\textwidth]{Figures/MatchedS10-2.png}} \\
%\subfigure[][$S=5a$ and $H=20a$]{\includegraphics[width=0.35\textwidth]{Figures/MatchedH20S5.png}}
%\subfigure[][$S=15a$ and $H=5a$]{\includegraphics[width=0.35\textwidth]{Figures/MatchedH5S15.png}} 
%\caption{Comparison of the asymptotically matched and exact solutions for $\sigma_{xy}$ as a function of position ahead of the lower-left crack for (a) $S=5a$ and $H=10a$ and (b) $S=10a$ and $H=10a$ (c) $S=5a$ and $H=20a$ (d) $S=15a$ and $H=5a$. The dashed vertical line shows the matching point $l$ as presented in Eq. \ref{eq:optequ}}\label{fig:MatchedH10}
%\end{figure*} 
%
 With these approximate values of $k_{II}$ and $b_{II}$ along with the single-crack solutions for $T$, $b_I$ and $k_I$, we can apply Eq.\ref{eq:alphas} to obtain the crack-path, $\lambda(x)$.  This constitutes `Method III' for EP crack path determination.
 
\section{Results for Growth Paths}
As an example, the results for all three methods under different crack placements are shown in Fig. \ref{fig:Scompare}. In these graphs, $\alpha,\ \beta$, and $\gamma$ are plotted versus $S={r_0}\sin(\theta_0)$ or $H={r_0} \cos(\theta_0)$ for extension length $L_{stop}=0.1 a$ and $\sigma_x^{\infty}=0$.

The results in Fig. \ref{fig:Scompare} show that the second and the third methods agree relatively well with each other and with respect to the first method, which is our most precise one.
The effect of $\sigma^{\infty}_x$ on crack opening path is presented in Fig. \ref{fig:TsBetGam}. These plots show the behavior of $\beta$ and $\gamma$ when changing the stress biaxiality, $\sigma_x^{\infty}/\sigma_y^{\infty}$, for two initial crack configurations: $H=r_0 \cos(\theta_0)=15$ and $S=r_0\sin(\theta_0)=5$ in which the crack initially kinks downward, and $H=r_0 \cos(\theta_0)=10$ and $S=r_0\sin(\theta_0)=10$ in which the crack initially kinks upward.  We note that $\alpha$ is independent of $\sigma_{x}^{\infty}$. %\begin{figure*}
%\subfigure[][]{\includegraphics[width=0.32\textwidth]{Figures/Ts-Betta.png}\\
%             \includegraphics[width=0.32\textwidth]{Figures/Ts-Gamma.png}\\
%             \includegraphics[width=0.32\textwidth]{Figures/TStress.png}}\\
%\subfigure[][]{\includegraphics[width=0.32\textwidth]{Figures/Ts-BettaH10.png}\\
%             \includegraphics[width=0.32\textwidth]{Figures/Ts-GammaH10.png}\\
%             \includegraphics[width=0.32\textwidth]{Figures/TStressH10.png}}\\
%\caption{Behavior of $\beta$, $\gamma$, and the lower left crack's propagation path for various $\frac{\sigma_x^{\infty}}{\sigma_x^{\infty}}$. Initial crack placements are (a) $H=15a$ and $S=5a$ and (b) $H=10a$ and $S=10a$.}
%          \label{fig:TsBetGam}
%          \vspace{-10pt}
%\end{figure*}
Fig. \ref{fig:TsBetGam} also shows the path, drawn for the lower left crack,
in these cases. The lines, plotted for three different $\frac{\sigma_x^{\infty}}{\sigma_y^{\infty}}$, show the corresponding crack opening paths. The two plots are scaled identically, but with vertical amplification to ease discernment of the various curves.

  \rcom{ We provide a bit of intuition for these results. For a single crack, if $\sigma^{\infty}_x=0$, the stress near the crack tip actually has a non-zero $T$ stress given by $T=-\sigma^{\infty}_y$.  If $\sigma_x^{\infty}/\sigma_y^{\infty}>1$, the positive T stress means horizontal tension exists near the crack that biases the crack toward curving up after an initial upward kink.   If $\sigma_x^{\infty}/\sigma_y^{\infty}<1$, the compressive $T$ stress works to stabilize the path and minimize curvature after a kink.  In the two-crack case, a similar argument explains why $\beta$ changes signs at $\sigma_x^{\infty}/\sigma_y^{\infty}=1$ and why larger lateral tension tends to increase crack curvature in Fig \ref{fig:TsBetGam}.}

\section{Conclusion}
We have performed an analytical study of EP cracks, culminating in three different methods for predicting EP crack paths. The first method is the most robust, producing a solution based on a stress field that continually modifies as the cracks open. This method utilizes a perturbation analysis to first order in the stress field, treating crack deflection as the small parameter. The second method is a simpler method derived from the first under the assumption that the length of the extension is much smaller than the distance between the cracks.  In this case the solution is similar to the case that just one of the cracks propagates while the other is static. The third method is based on the assumption that the crack length is also smaller than the distance between two cracks. The third method works based on a superposition of the stresses from two isolated cracks, and matching this field to the asymptotic form of a Williams expansion near the crack tips.  Through comparison to an exact  stress solution for straight cracks, we provide a verification as to the correctness of predictions of Methods I and II, since exact solutions for full EP crack paths are not available. The results show a good mutual agreement between all three methods when the corresponding assumptions are valid. 
The validity of Method III opens many possible doors for modeling of systems with larger numbers of interacting cracks.  As the number of cracks increases, the complexity of the integral system that represents the full solution grows significantly.  However, the described procedure of Method III remains at its core quite simple; one builds an outer solution as a superposition of single-crack fields as if each crack were on its own, and then applies the matching argument at all cracks to approximate the needed Williams expansion coefficients at each crack tip.  However, some subtleties must be addressed before this notion may be applied to a many-crack system.  For instance, our perturbation method assumes a small deflection of the crack tips, which is appropriate for EP crack when they are not too close to each other, since the mode-mixity on each crack is low.  However, in a system with an arbitrary scattering of cracks, large initial kink angles may occur, which might force the cracks to propagate in a fashion contrary to this assumption.  Furthermore, in the EP system, it could be assumed that both cracks open symmetrically, however in a general system of many cracks, $k_I$ varies from crack to crack and one would have to track which cracks are critical as the far-field loading is increased in order to model the correct progression of propagation. Crack growth aside, the third method is essentially equivalent to an approximation technique for Williams expansion coefficients, which could have value in its own right for approximating stress intensity factors and higher order terms for a general crack geometry.  We leave exploration on this front as future research.
\section{Acknowledgement}
R. Ghelichi acknowledges support from the MIT-Italy program through the Progetto Rocca fellowship and from the MIT Department of Mechanical Engineering. K. Kamrin acknowledges support from the MIT Department of Mechanical Engineering. The authors thank David Parks for helpful comments and advice.  
\clearpage

\newpage
\appendix
\newpage
\section{Truncation in Williams expansion}
\label{app:kIext}
\begin{figure}
\begin{center}
\includegraphics[width=0.4\textwidth]{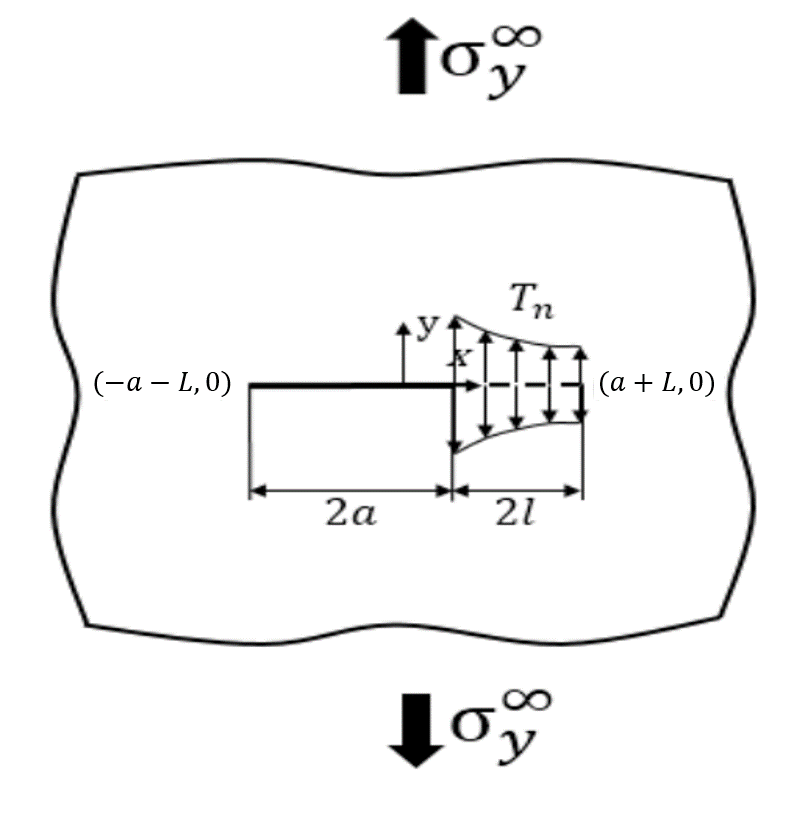}
\caption{A Griffith crack with the length of $2a$ under tension has been extended from one side to the length of $2l$}
\label{fig:kIext}
\end{center}
\end{figure}

\begin{figure}
\begin{center}
\includegraphics[width=0.45\textwidth]{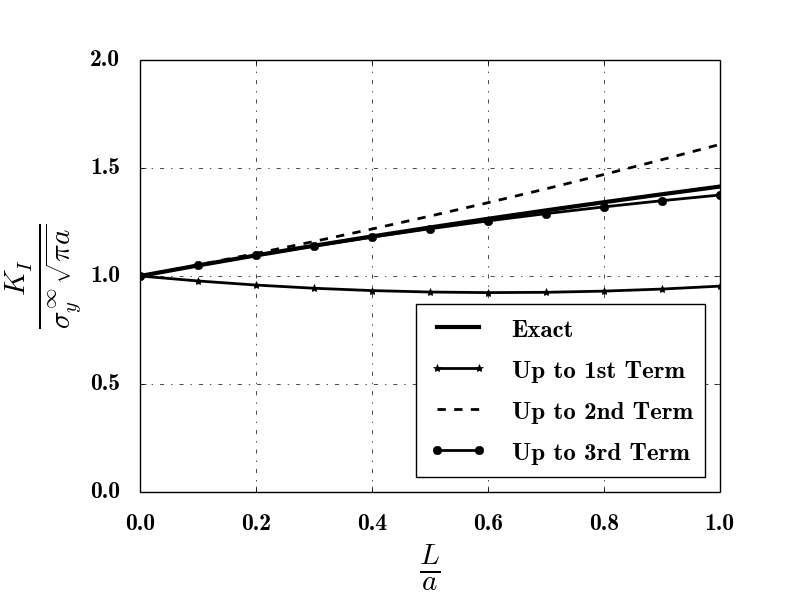}
\caption{Comparison against the exact solution of $K_I$ for an extended crack as obtained by keeping different numbers of terms in Williams expansion of the original crack using, Eq. \ref{eq:mustrac}.}
\label{fig:kIpert}
\end{center}
\end{figure}

\begin{figure*}
\subfigure[][$H/a=1$ and $S/a=1$]{\includegraphics[width=0.3\textwidth]{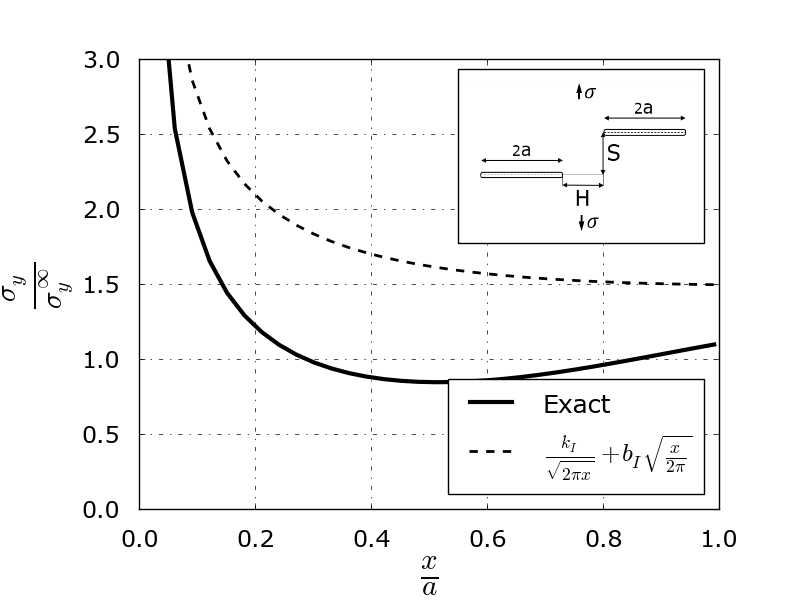}} 
\subfigure[][$H/a=3$ and $S/a=1$]{\includegraphics[width=0.3\textwidth]{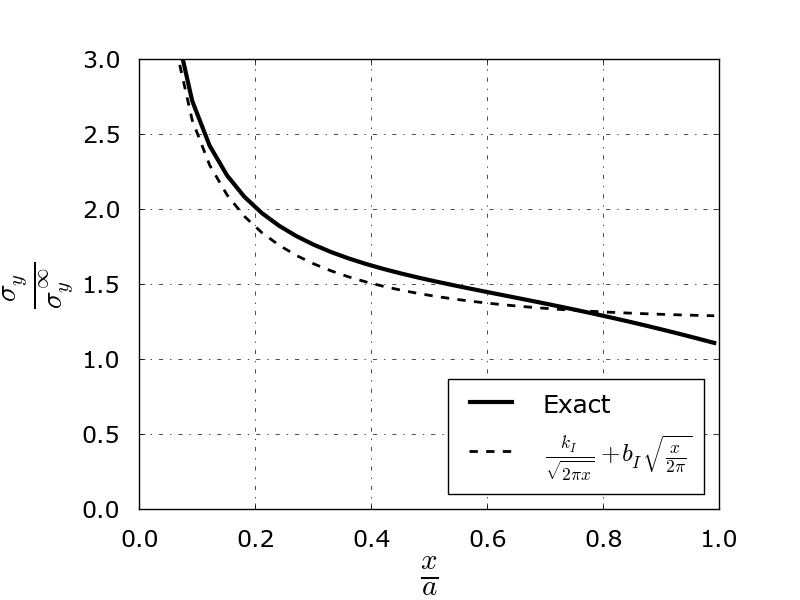}}
\subfigure[][$H/a=5$ and $S/a=1$]{\includegraphics[width=0.3\textwidth]{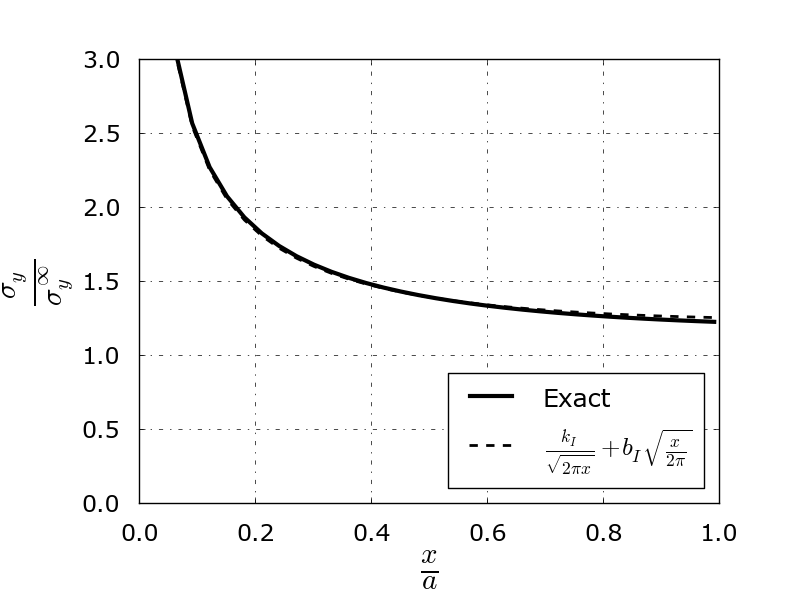}}
\caption{The stress near the crack tip as given by the first two terms of the Williams expansion is compared to the exact solution in several EP geometries.}
\label{fig:CFeval}
\end{figure*}

To study how truncating the close-field Williams expansion affects the accuracy of our method, we first consider the extension of a single Griffith crack with length of $2a$ as presented in Fig. \ref{fig:kIext}.  We use our integral method to extend the crack from its right side by a length of $2l$. For the close-field solution we use stresses that keep a differing number of Williams expansion terms.  Because we also have an exact solution for the extended crack, we can compare the different predictions to examine the effect of truncating the Williams expansion after a finite number of terms.
Applying the integral method from Sec \ref{eq:fandw} to approximate the stress field post-extension, the value of $K_I$ of the extended crack can be obtained exactly from the formula
\begin{equation}
K_I=\frac{1}{\sqrt{\pi (a+L)}}\int_{-a-L}^{+a+L} T_n(x) \sqrt{\frac{a+L+x}{a+L-x}} dx
\label{eq:kIext}
\end{equation}
where $T_n$, as presented in Fig. \ref{fig:kIext}, is the close-field solution given from the unextended crack, which is defined by  
\begin{equation}
\begin{aligned}
T_n &= \begin{cases} 0 &\mbox{if } -a-L \leq x < a-L \\ 
\widetilde{T}_n(x) & \mbox{if } x \geq a-L. \end{cases} 
\label{eq:mustrac}
\end{aligned}
\end{equation}
$\widetilde{T}_n(x)$ has an exact solution 
\begin{equation}
\widetilde{T}_n(x)=\frac{\sigma_y^{\infty} (x+L)}{\sqrt{(x+L)^2-a^2}}, \  x \geq a-L 
\label{eq:Texact}
\end{equation}
or can be approximated in the form of a truncated Williams expansion,
\begin{equation}
\begin{aligned}
\widetilde{T}_n(x)&=\frac{\sigma_y^{\infty}\sqrt{a}}{\sqrt{2(x-a+L)}} + \frac{3 \sigma_y^{\infty} \sqrt{-a+L+x}}{4 \sqrt{2a}}-\frac{5 \sigma_y^{\infty} (-a+L+x)^{3/2}}{32 \sqrt{2} a^{3/2}}+\\
&+O(x-a+L)^{5/2}.
\end{aligned}
\label{eq:Twill}
\end{equation}
From the exact $\widetilde{T}_n(x)$, Eq \ref{eq:kIext} gives the correct value of $K_I=\sigma_y^{\infty} \sqrt{\pi (a+L)}$.  Fig. \ref{fig:kIpert} shows the different formulae for $K_I(L)$ obtained by keeping different numbers of terms in the Williams expansion.  Fig. \ref{fig:kIpert} shows that the singular term in Williams expansion is not enough to capture even the first order variation of the exact results after extension.  In fact it appears that with every additional term kept, resulting $K_{I}(L)$ gains another derivative of accuracy at $L=0$. This point does not appear to have been acknowledged in existing literature.  Our decision to truncate the close-field Williams expansions in Eq \ref{eq:asymStress} at the square-root term reflects our desire to represent the stresses (and stress intensity) after extension at a reasonable accuracy, capturing the first-order variations in stress intensity as the crack grows.
Still, any truncation limits the region of accuracy of the close-field solution, which limits the maximum length $L_{stop}$ to which the crack may be accurately extended. The size of the region of accuracy is somewhat problem-dependent.  Beside the iterative approach suggested in the main text, another solution to this issue would be to use a more accurate close-field solution; i.e. keeping more expansion terms or otherwise finding a better initial stress field in the analysis.  Recall that the close-field solution is a reference solution assumed to be given a priori, and our analysis can be applied to any such reference solution.  Figure \ref{fig:CFeval}(a-c) compares the asymptotic solution obtained from the first two terms of the Williams expansion to the exact stress field for different systems of EP cracks. 
The latter graph shows the accuracy of the results depends on the positions of the two cracks. For $H=1$ and $S=1$ the Williams expansion quickly loses accuracy away from the crack tips, while for $H=5$ and $S=1$ the solution stays in good agreement for a much longer distance from the crack tips. 
%
%\clearpage
\section{Analysis of Method III}\label{app:C}
\begin{figure}[t]
\centering
\subfigure[][$\frac{Re(\phi^{res}(z_1))}{\sigma_y^{\infty}}$]{\includegraphics[width=0.45\textwidth]{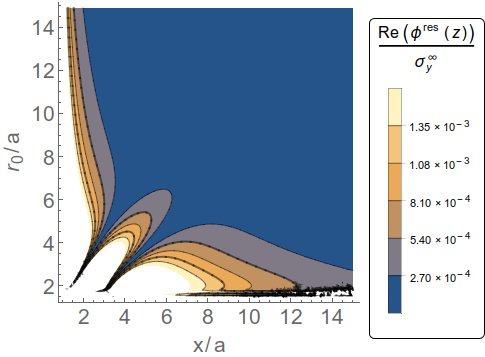}}
\subfigure[][$\frac{Re(\phi_{Est}^{res}(z_1))}{\sigma_y^{\infty}}$]{\includegraphics[width=0.45\textwidth]{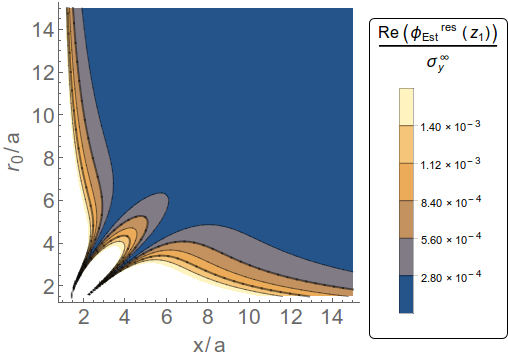}}\\
\caption{Behavior  of a) $\frac{Re(\phi^{res}(z_1))}{\sigma_y^{\infty}}$ b) $\frac{Re(\phi_{Est}^{res}(z_1))}{\sigma_y^{\infty}}$ 
as it varies with $r_0$ and position, $x$, ahead of the lower-left crack, i.e. $x=Re(z_1)$ and $Im(z_1)=0$ for cracks offset by an angle $\theta_0=\pi/6$.}
\label{fig:phires}
\end{figure}

The purpose of this appendix is to provide analytical backing to the matching technique we call Method III.  The method hinges on the existence of an overlapping zone of accuracy shared by an outer solution, which we take to be the solution of two superposed cracks, and an inner solution, which we express as a two-term Williams expansion.  Herein, we derive an error formula for the outer and inner solutions and derive criteria for when the two approximations should be simultaneously valid within a shared window of accuracy.
The solution for a system of two cracks can be expressed as the sum of the solutions for each crack per Eq \ref{sumcrackstress} added to a residual stress coming from the effect of crack interaction. Letting $\phi^{sum}$ represent the Muskelishvili potential for the sum of the two individual crack solutions and $\phi^{res}$ be the potential for the residual field, we have
\begin{equation}
\begin{aligned}
&\phi(z_1)=\phi^{sum}(z_1)+\phi^{res}(z_1), \ \phi^{sum}(z_1)=\phi^G(z_1)+\phi^G(z_0-z_1). \\
\end{aligned}
\label{eq:phires}
\end{equation}
The complete Muskhilishvili's potential for two parallel offset cracks, $\phi(z_1)$,  can be written in non-closed form from the solution in Horii and Nemat-Nasser \cite{hori1985}.  Let  $\zeta\equiv \frac{a}{|r_c|}$, in which $r_c=r_0e^{i \theta_0}+2a=|r_c|e^{\theta_c}$ is the line connecting the center of the two cracks.  Upon expanding the complete solution, one obtains the following relation
\begin{equation}
\begin{aligned}
\phi(z_1)=&\phi^1(z_1)+\phi^2(z_0-z_1)=\phi^{sum}(z_1)+\underbrace{\phi^{sum}(z_1)f(\zeta,\theta_c)}_\text{$\equiv \phi_{Est}^{res}(z_1)$}+O\left(\frac{\zeta^3}{\sqrt{z_1}}\right) \\
\end{aligned}
\label{eq:Phresest}
\end{equation}
where
\begin{equation}\label{fandphi12}
\begin{aligned}
f(\zeta,\theta_c)&=\zeta^2((\cos2 \theta_0-\frac{1}{2}\cos4 \theta_c)+ i\frac{1}{2}(\sin2 \theta_c - \sin4 \theta_c)) \\
\phi^j(z_j)&=\underbrace{\phi^G(z_j)+\phi^G(z_j)f(\zeta,\theta_c)}_\textbf{$\equiv \phi_{Est}^j(z_j)$}+O\left(\frac{\zeta^3}{\sqrt{z_j}}\right); \ \ \ j=1,2. \\
\end{aligned}
\end{equation}
Equation \ref{eq:Phresest} gives a formula for $\phi_{Est}^{res}$, the leading-order approximation for $\phi^{res}$. To further demonstrate the accuracy of our $\phi_{Est}^{res}$ formula, Fig. \ref{fig:phires} shows the exact $\phi^{res}(z_1)$ along with $\phi_{Est}^{res}(z_1)$ presented in Eq. \ref{eq:Phresest}, as distance between the cracks, $r_0$, is varied at constant $\theta_0=\frac{\pi}{6}$.  We can conclude that the difference between the exact $\phi$ and $\phi^{sum}$ decays as $\phi_{Est}^{res}$, which behaves $\sim\zeta^2/z_{1}^2$, confirming that the true solution always approaches the outer solution as both distance and crack separation increase.
Eq.\ref{eq:Phresest} permits us to compute the precision of Method III with respect to the exact stress field. From the exact potential $\phi=\phi^1+\phi^2$, we can write the exact shear stress $\sigma_{xy}$ for the two-crack problem as
\begin{equation}
\begin{aligned}
\sigma_{xy}(z_1)&=\sigma^1_{xy}(z_1)+\sigma^2_{xy}(z_0-z_1).
\end{aligned}
\label{eq:sigxysup}
\end{equation}
These component fields can be expanded per Eq \ref{fandphi12} as
\begin{equation}\label{sigma12}
\begin{aligned}
\sigma^j_{xy}(z_j)&=-2Im(z_j)Re(\phi_{Est}'^{j}(z_j))-2Im(f(\zeta,\theta_c))Im(\phi^G(z_j))+\\
&\hspace{1cm} +O\left(\frac{\zeta^3}{\sqrt{z_j}}\right) \ \ j=1,2
\end{aligned}
\end{equation}
For $\sigma^1_{xy}$ the above simplifies to
\begin{equation}
\begin{aligned}
\sigma^1_{xy}(x_1) &=-2Im(f(\zeta,\theta_c)) \phi^G(x_1)+O\left(\frac{\zeta^3}{\sqrt{x_1}}\right).
\end{aligned}
\label{eq:sig1}
\end{equation}
Let the expansion of the Griffith's crack solution about the crack tip be denoted
\begin{equation}
\phi^G(x_1) = -\frac{\sigma_y^{\infty}}{2}+\sum_{n=-1}^{\infty}a_n x_1^{1/2+n}.
\end{equation}
Similarly we can obtain
\begin{equation}
\begin{aligned}
\sigma^2_{xy}(z_0-x_1) &= \overbrace{-2Im(z_0)Re(\phi'^{G}(z_0-x_1))}^\text{$\sigma_{xy(outer)}(x_1)$}-\\
&-2Im(z_0)Re(f(\zeta,\theta_c)\phi'^{G}(z_0-x_1))\\
&-2Im(f(\zeta,\theta_c)) Re(\phi^G(z_0-x_1))+O\left(\frac{\zeta^3}{\sqrt{x_1}}\right) \\
\end{aligned}.
\label{eq:sig2}
\end{equation}
where we have noted the emergence of the  outer solution, Eq \ref{sumcrackstress}.  Let us define the Taylor expansion
\begin{align}\label{taylor_alpha}
 \sum_{n=0}^{\infty} \alpha_n x_1^n\equiv& -2Im(z_0)Re(\phi'^{G}(z_0-x_1))-2Im(z_0)Re(f(\zeta,\theta_c)\phi'^{G}(z_0-x_1)) \nonumber
\\
&-2Im(f(\zeta,\theta_c)) Re(\phi^G(z_0-x_1)).
\end{align}
Near $z_1=0$ the exact shear stress $\sigma_{xy}$ has the following form of Williams expansion
\begin{equation}
\sigma_{xy}(x_1)=\frac{k_{II}}{\sqrt{2 \pi x_1}}+b_{II} \sqrt{\frac{x_1}{2 \pi}}+c_{II} \frac{x_1}{\sqrt{2 \pi}} + O(x_1^{3/2})
\label{eq:willsigxy}
\end{equation}
ahead of the crack tip. By combining Eq.\ref{eq:willsigxy}, Eq. \ref{sigma12} and Eq.\ref{eq:sigxysup} the following results can be obtained:
\begin{equation}\label{asympk}
\begin{aligned}
k_{II}&=-2\sqrt{2 \pi} Im(f(\zeta,\theta_c)) a_{-1}+O(\zeta^3)= \\
&\frac{\sigma_y^{\infty}\sqrt{\pi a}}{2}\zeta ^2(\sin4\theta_c-\sin2\theta_c)+O(\zeta^3)\\
b_{II}&=-2\sqrt{2 \pi} Im(f(\zeta,\theta_c)) a_{0}+O(\zeta^3)=\\
&\frac{\sigma_y^{\infty} 3 \sqrt{\pi}}{8\sqrt{a}}\zeta ^2(\sin4\theta_c-\sin2\theta_c)+O(\zeta^3)\\\
c_{II}&=\sqrt{2 \pi}\alpha_1 +O(\zeta^3)\\
\end{aligned}
\end{equation}
It bears noting that the above formulas for $k_{II}$ and $b_{II}$ are useful on their own as strong approximations for the shear stress intensity factors of EP cracks.  These could be used to approximate $\alpha$, $\beta$, and $\gamma$ for the crack path, however, they are specific to the EP geometry and do not share the apparent generality that we may hope to gain in the future from the matching approach of Method III.
In Sec \ref{matching} the inner solution is defined as a two-term Williams expansion based on a functional matching with the outer solution at the matching point.  As will be discussed more in a moment, for the purposes of backward error evaluation in Method III, we assume for now an inner solution with exact coefficients,
\begin{equation}
\sigma_{xy(inner)} \cong \frac{k_{II}}{\sqrt{2 \pi x_1}}+b_{II} \sqrt{\frac{x_1}{2 \pi}}.
\end{equation}
We define the error of the inner/outer solutions by 
\begin{equation}
\begin{aligned}
e_{inner}(x_1)&=\sigma_{xy}(x_1)-\sigma_{xy(inner)}(x_1)\\
e_{outer}(x_1)&=\sigma_{xy}(x_1)-\sigma_{xy(outer)}(x_1)\\
\end{aligned}
\label{eq:ees}
\end{equation}
By combining the Eq.\ref{eq:ees} with Eq.\ref{eq:sig1} and Eq.\ref{eq:sig2}, we obtain the following form for the inner solution error,
\begin{equation}
\begin{aligned}
e_{inner}(x_1) &= \sigma^1_{xy}(x_1)+\sigma^2_{xy}(x_1)-\sigma_{xy(inner)}(x_1)\\
e_{inner}(x_1) &= c_{II}\frac{ x_1}{\sqrt{2\pi}}+O(x_1^{3/2})\\
e_{inner}(x_1) &= \underbrace{\alpha_1x_1}_{\equiv e_{inner}^{Est}(x_1)} +O(x_1 \zeta^3) +O(x_1^{3/2})
%\zeta^2 c_1 &= -2 Im(z_0)Re\bigg(f(\zeta,\theta_c)\frac{3 \sigma_y^{\infty}a^2(a+z_0)}{2z_0^{5/2}(2a+z_0)^{5/2}}\bigg)\\
%&\hspace{0.5in}- Im(f(\zeta,\theta_c))Re\bigg(\frac{a^2 \sigma_y^{\infty}}{2z_0^{3/2}(2a+z_0)^{3/2}}\bigg)\\
%e_{inner(Est)}(x_1) & = (b_1 + \zeta^2 c_1) x_1 
\end{aligned}\label{eq:Einner}
\end{equation}
where, from Eq \ref{taylor_alpha}, we calculate
\begin{align}
\alpha_1 =&  \frac{3 a^2 \sigma_y^{\infty} Im(z_0) (a - z_0)}{(z_0 (-2 a + z_0))^{5/2}} -2 Im(z_0)Re\bigg(f(\zeta,\theta_c)\frac{3 \sigma_y^{\infty}a^2(a+z_0)}{2z_0^{5/2}(2a+z_0)^{5/2}}\bigg)\\
&\hspace{0.5in}- Im(f(\zeta,\theta_c))Re\bigg(\frac{a^2 \sigma_y^{\infty}}{2z_0^{3/2}(2a+z_0)^{3/2}}\bigg).
\end{align}
As for the outer solution error, we have
\begin{equation}
\begin{aligned}
e_{outer}(x_1) &= \sigma^1_{xy}(x_1)+\sigma^2_{xy}(x_1)-\sigma_{xy(outer)}(x_1)\\
e_{outer}(x_1) &= -2Im(f(\zeta,\theta_c)) \phi^G(x_1)-2Im(z_0)Re(f(\zeta,\theta_c)\phi'^{G}(z_0-x_1))-\\
&\hspace{0.5in} -2Im(f(\zeta,\theta_c)) Im(\phi^G(z_0-x_1))+O(\frac{\zeta^3}{\sqrt{x_1}}) \\
 e_{outer}^{Est}(x_1) &\equiv -2Im(f(\zeta,\theta_c)) \phi^G(x_1)-2Im(z_0)Re(f(\zeta,\theta_c)\phi'^{G}(z_0-x_1))-\\
&\hspace{0.5in} -2Im(f(\zeta,\theta_c)) Im(\phi^G(z_0-x_1)).
\end{aligned}
\label{eq:Eouter}
\end{equation}
\begin{figure}
\centering
\subfigure[][$S=5a$ and $H=15a$]{\includegraphics[width=0.4\textwidth]{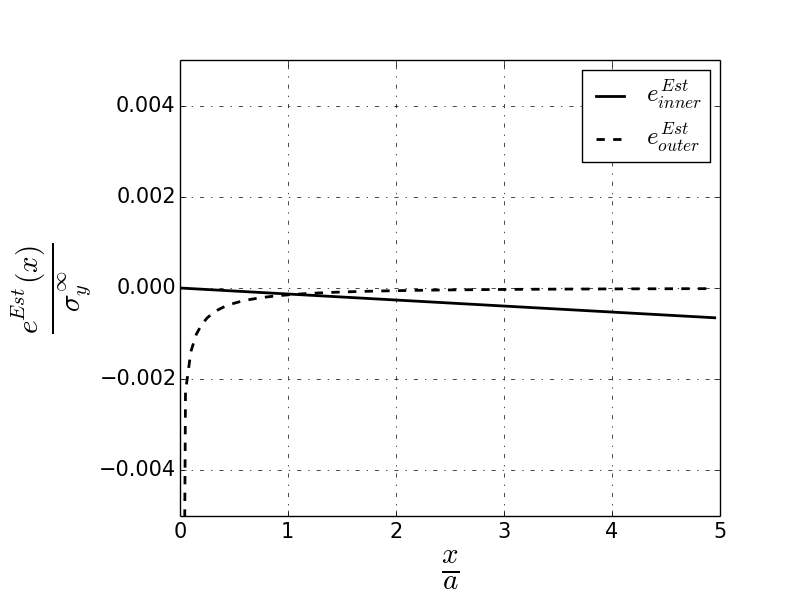}}
\subfigure[][$S=1a$ and $H=5a$]{\includegraphics[width=0.4\textwidth]{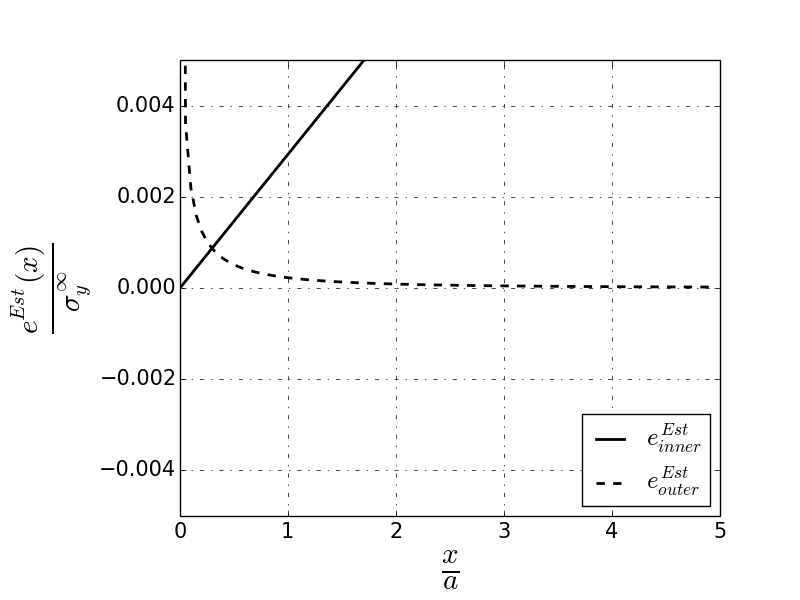}} \\
\caption{Behavior of $e_{inner}$ and $e_{outer}$ in two cases (a) $S=5a$ and $H=15a$ and (b) $S=1a$ and $H=5a$.}
\label{fig:Error}
\end{figure}
In Method III, we would like to split the domain for $\sigma_{xy}(z_1)$ in the following form:
\begin{equation}
\sigma_{xy}(x_1)\cong \begin{cases}
   \sigma_{xy(inner)}(x_1) & \text{if } x_1 \leq l \\
   \sigma_{xy(outer)}(x_1) & \text{if } x_1 \geq l
  \end{cases} 
\label{eq:sxysplit} 
\end{equation}
with the goal of $l$ representing a representative location where both inner and outer solutions agree sufficiently with the exact solution.  Thus, Method III is expected to be valid within some tolerance $\epsilon$ when there exists an $l$ such that 
\begin{equation}\label{accuracy_crit}
\text{max}(|e^{Est}_{inner}(l)|,|e^{Est}_{outer}(l)|, |e'^{Est}_{inner}(l)|,|e'^{Est}_{outer}(l)|,|e''^{Est}_{inner}(l)|,|e''^{Est}_{outer}(l)|)<\epsilon.
\end{equation}
To be clear, we remind that the way Method III is actually used requires one to first construct the outer solution and then build the inner solution from the outer solution at a matching point.  Rather than measure the error of the constructed inner solution, the criterion above measures the backward error by assuming an inner solution with correct $k_{II}$ and $b_{II}$, composing it with the outer solution at various $l$, and measuring how much the composite function would disagree with the exact solution at $l$.  For example, if (\ref{accuracy_crit}) were satisfied for $\epsilon=0$ at some $l$, then the inner solution we would construct from the outer solution at $l$ would give the exact $k_{II}$ and $b_{II}$. The above measurement of error both reflects the accuracy of Method III, and, as described here, is tractable to write for analytical error analysis.

The behavior of $e^{Est}_{inner}$ and $e^{Est}_{outer}$ are shown in two examples presented in Fig.\ref{fig:Error}.  Per (\ref{accuracy_crit}), we would expect the validity of Method III to be stronger for the case shown in Fig.\ref{fig:Error}(a) compared to that of Fig.\ref{fig:Error}(b), and this expectation is strongly confirmed referring back to Fig. \ref{fig:Scompare}.

\section{Series Coefficients}
\label{app:A}
The constants in Eq.\ref{eq:fandw} based on the asymptotic solution for the two parallel cracks. \\
%\begin{widetext}
\begin{align*}
F_{Qn} = f_{i\frac{n}{2}}(\alpha L n &+ 2\alpha)+f_{i\frac{n+1}{2}}(n \beta L^{3/2}+\frac{3}{2}\beta L^{1/2})+f_{i\frac{n+2}{2}}(n\gamma L^2+4 \gamma L)
\\
 &-(\alpha L+\beta L^{3/2}+\gamma L^2)f_{i\frac{n-2}{2}}
 \\
F_{Pn}=n(\alpha L f_{i\frac{n}{2}}&+\beta L^{3/2} f_{i\frac{n+1}{2}}+\gamma L^2 f_{i\frac{n+2}{2}}-(\alpha L+\beta L^{3/2}+\gamma L^2)f_{i\frac{n-2}{2}})\\
f_{in}=&-\frac{(-1)^i \Gamma (n+1)}{(2 i+2 n+3) \Gamma \left(-i-\frac{1}{2}\right) \Gamma (i+1) \Gamma \left(n+\frac{3}{2}\right)}\\
v_{in}=&\frac{2 (-1)^i L^{n-1} \Gamma (n+1)}{(2 i+2 n+1) \Gamma \left(-i-\frac{1}{2}\right) \Gamma (i+1) \Gamma \left(n+\frac{1}{2}\right)}
\end{align*}

$\Gamma(t)$ is the Gamma function, $\Gamma(t)=\int_0^{\infty}x^{t-1}e^{-x}dx$.
\begin{align*}
A_{-1/2}&=\left[\left(k_I-\frac{3}{2}\alpha k_{II}\right)-i\left(k_{II}+ \frac{\alpha}{2}k_I\right)\right]\frac{1}{\sqrt{2\pi L}}\\
A_{0}&=\left[-\frac{5\beta k_{II}}{2\sqrt{2\pi}}-i\left(\alpha T+\frac{\beta k_I}{2\sqrt{2\pi}}\right)\right]\\
A_{1/2}&=\left[ \left(b_I-\frac{7}{2}\gamma k_{II}-\frac{5}{2}\alpha b_{II} \right)-i\bigg(b_{II}-3\sqrt{\frac{\pi}{2}}\beta T+\right.\\
&\left.+\frac{\gamma k_I}{2}-\frac{\alpha b_I}{2}\bigg)\right]\sqrt{\frac{L}{2\pi}}\\
\end{align*}
\begin{align*}
B_{-1/2}&=\left[\left(\alpha^2 k_I + 2\alpha k_{II}\right)+i \left(2\alpha k_I - 3\alpha^2 k_{II}\right)\right]\frac{1}{\sqrt{2\pi L}}\\
B_{0}&=\frac{5 \alpha  \beta  k_I}{2 \sqrt{2 \pi }}+\frac{3 \beta  k_{II}}{\sqrt{2 \pi }}-2 \alpha ^2 T +i \left(\frac{3 \beta k_I}{\sqrt{2 \pi }}-\frac{19 \alpha  \beta  k_{II}}{2 \sqrt{2 \pi }}\right)\\
B_{1/2}&=\sqrt{L} \left(-\frac{\alpha ^2 b_I}{\sqrt{2 \pi }}+\sqrt{\frac{2}{\pi }} \alpha  b_{II}+\frac{3 \alpha  \gamma  k_I}{\sqrt{2 \pi }}+\right.\\
&+\left.\frac{3 \beta ^2 k_I}{2 \sqrt{2 \pi }}+2 \sqrt{\frac{2}{\pi }} \gamma  k_{II}-6 \alpha  \beta  T\right)+\\
& i \sqrt{L} \left(\sqrt{\frac{2}{\pi }} \alpha  b_I-\frac{5 \alpha ^2 b_{II}}{\sqrt{2 \pi }}+2 \sqrt{\frac{2}{\pi }} \gamma  k_I-\frac{13 \alpha  \gamma  k_{II}}{\sqrt{2 \pi }}-\frac{15 \beta ^2 k_{II}}{2 \sqrt{2 \pi }}\right)\\
B_{1}&=L \left(-\frac{3 \alpha  \beta  b_I}{2 \sqrt{2 \pi }}+\frac{3 \beta  b_{II}}{\sqrt{2 \pi }}+\frac{7 \beta  \gamma  k_I}{2 \sqrt{2 \pi }}-4 \alpha  \gamma  T-\frac{1}{2} 9 \beta ^2 T\right)+\\
&+\frac{i \beta  L (6 b_I-15 \alpha  b_{II}-41 \gamma  k_{II})}{2 \sqrt{2 \pi }}\\
B_{3/2}&=\gamma \sqrt{\frac{2}{\pi }} L^{3/2} \left(- \alpha  b_I+2 b_{II}+ \gamma  k_I-6 \beta  T\right)+\\
&+i \sqrt{\frac{2}{\pi }} \gamma  L^{3/2} (2 b_I-5 \alpha  b_{II}-7 \gamma  k_{II})\\
\end{align*}

%\end{widetext}
%%\end{equation*}
%%\begin{equation*}
%\clearpage
%\begin{widetext}
\begin{align*}
C_{-1/2}&=-\frac{L (\alpha  k_I+2 k_{II}) \left(\alpha +\beta  \sqrt{L}+\gamma  L\right)}{2 \sqrt{2 \pi }}-\\
&-\frac{i L (2 k_I-3 \alpha  k_{II}) \left(\alpha +\beta  \sqrt{L}+\gamma  L\right)}{2 \sqrt{2 \pi }}\\
C_{0}&=\frac{L \left(4 \sqrt{\pi } \alpha  T-\sqrt{2} \beta  k_I\right) \left(\alpha +\beta  \sqrt{L}+\gamma  L\right)}{4 \sqrt{\pi }}+\\
&+\frac{5 i \beta  k_{II} L \left(\alpha +\beta  \sqrt{L}+\gamma  L\right)}{2 \sqrt{2 \pi }}\\
C_{1/2}&=\frac{1}{4 \sqrt{\pi }}(\sqrt{2} \alpha  \beta  b_I L^{3/2}+\sqrt{2} \alpha  b_I \gamma  L^2+\sqrt{2} \alpha ^2 b_I L-\\
&-2 \sqrt{2} b_{II} L \left(\alpha +\beta  \sqrt{L}+\gamma  L\right)+\sqrt{2} \alpha ^2 k_I-\\
&-\sqrt{2} \beta  \gamma  k_I L^{3/2}-\sqrt{2} \gamma ^2 k_I L^2-\sqrt{2} \alpha  \gamma  k_I L+2 \sqrt{2} \alpha  k_{II}+\\
&+6 \sqrt{\pi } \beta ^2 L^{3/2} T+6 \sqrt{\pi } \beta  \gamma  L^2 T +6 \sqrt{\pi } \alpha  \beta  L T)+\\
&+i \frac{1}{2 \sqrt{2 \pi }}(-2 b_I L \left(\alpha +\beta  \sqrt{L}+\gamma  L\right)+5 \alpha  \beta  b_{II} L^{3/2}+5 \alpha  b_{II} \gamma  L^2+\\
&+5 \alpha ^2 b_{II} L+2 \alpha  k_I-3 \alpha ^2 k_{II}+7 \beta  \gamma  k_{II} L^{3/2}+7 \gamma ^2 k_{II} L^2+7 \alpha  \gamma  k_{II} L)\\
\end{align*}
\begin{align*}
C_{1}&=\frac{\beta  (\alpha  k_I+k_{II})}{\sqrt{2 \pi }}+\frac{i \beta  (k_I-4 \alpha  k_{II})}{\sqrt{2 \pi }}+\alpha ^2 (-T)\\
\end{align*}
\begin{align*}
C_{3/2}&=\frac{1}{4 \sqrt{\pi }}\bigg(-\sqrt{2} \alpha ^2 b_I+2 \sqrt{2} \alpha  b_{II}+2 \sqrt{2} \alpha  \gamma  k_I+\\
&+\sqrt{2} \beta ^2 k_I+2 \sqrt{2} \gamma  k_{II}-10 \sqrt{\pi } \alpha  \beta  T\bigg)+\\
&+\frac{i \left(2 \alpha  b_I-5 \alpha ^2 b_{II}+2 \gamma  k_I-10 \alpha  \gamma  k_{II}-5 \beta ^2 k_{II}\right)}{2 \sqrt{2 \pi }}\\
\end{align*}
\begin{align*}
C_{2}&=-\frac{\alpha  \beta  b_I}{2 \sqrt{2 \pi }}+\frac{i \beta  (2 b_I-5 \alpha  b_{II}-12 \gamma  k_{II})}{2 \sqrt{2 \pi }}+\\
&+\frac{\beta  b_{II}}{\sqrt{2 \pi }}+\frac{\beta  \gamma  k_I}{\sqrt{2 \pi }}-\alpha  \gamma  T-\frac{1}{2} 3 \beta ^2 T\\
C_{5/2}&=\frac{\gamma  \left(-\sqrt{2} \alpha  b_I+2 \sqrt{2} b_{II}+\sqrt{2} \gamma  k_I-6 \sqrt{\pi } \beta  T\right)}{4 \sqrt{\pi }}-\\
&-\frac{i \gamma  (-2 b_I+5 \alpha  b_{II}+7 \gamma  k_{II})}{2 \sqrt{2 \pi }}\\
\end{align*}
\section{Calculation of $\phi_0^{1}$ and $\phi_1^{1}$}\label{app:B}
\begin{equation}
\begin{aligned}
\phi_0^1(z_1) =\Omega_0^1(z_1)&=\sum_{m=-1}^1 A_{\frac{m}{2}} \sum_{i=0}^{\infty}f_{i\frac{m}{2}}\left( \frac{L}{z_1}\right)^{i+3/2}+\\
&+\sum_{n=0}^{N} (P_{\frac{n}{2}}-i Q_{\frac{n}{2}})\sum_{i=0}^{\infty}f_{i\frac{n}{2}}\left( \frac{L}{z_1}\right)^{i+3/2} 
\\
\phi_1^1(z_1)+\Omega_1^1(z_1)&=\sum_{m=-1}^{3}B_{\frac{m}{2}}\sum_{i=0}^{\infty}f_{i\frac{m}{2}}\left(\frac{L}{z_1}\right)^{i+3/2}+\\
&+\sum_{m=-1}^{5}C_{\frac{m}{2}}\sum_{i=0}^{\infty}v_{i\frac{m}{2}}\left(\frac{L}{z_1}\right)^{i+3/2}
\\
&+\sum_{n=0}^{N}\sum_{i=0}^{\infty}(Q_{\frac{n}{2}} F_{Qn}-i  P_{\frac{n}{2}} F_{Pn})\left(\frac{L}{z_1}\right)^{i+3/2}
\\
\end{aligned}
\label{equ:field}
\end{equation}
where $N$ is the number of the terms one chooses to keep in the Taylor expansion in Eq.\ref{eq:mirrortraction}; the variables $f_{in}$, $F_{Pn}$, $F_{Qn}$, $A_k$, $B_k$, and $C_k$ can be obtained based on the Williams expansion coefficients, which are presented in Appendix \ref{app:A}.
By substituting the above into Eq \ref{eq:system}, the unknown constants in the Taylor series and crack path will be obtained through a solvable non-linear system of equations. The final equations for $P_n$ and $Q_n$ are expressed as follows:
\begin{equation}
\begin{aligned}
P_n&=Re(\phi^{*}_1(n))+2Re(\phi^{*}_0(n))+2 \alpha Im(\phi^{*}_0(n))(n+2)\\
&+\gamma Im(\phi^{*}_0(n-1))(n+1)\delta _{n-1}-\lambda(L)(n+1)Im(\phi^{*}_0(n+1))\delta _{N-n-1}\\
P_{\frac{n}{2}}&=2\beta n Im(\phi^{*}_0(n))+6\gamma Im(\phi^{*}_0(n))\\
Q_n&=Im(\phi^{*}_1(n))+2Im(\phi^{*}_0(n))+\alpha n Re(\phi^{*}_0(n))
+\gamma Re(\phi^{*}_0(n-1))n\delta _{n-1}+\\
&+\lambda(L)(n+1)Re(\phi^{*}_0(n+1))\delta _{N-n-1}\\
Q_{\frac{n}{2}}&=2\beta n Re(\phi^{*}_0(n))\\
&n=0,1,...,N;\ \ \ \ \delta_n=\begin{cases} 1 &\mbox{if } n \geq 0 \\ 
0 & \mbox{if } n < 0. \end{cases}
\end{aligned}
\label{eq:NEqu}
\end{equation}

The constants in Eq.\ref{eq:NEqu} are:\\
%\begin{widetext}
\begin{align*}
Re(\phi^{*}_1(n))&=\sum_{m=-1}^{3}Re(B_{\frac{m}{2}})Re(HF_{\frac{m}{2}n})+\sum_{m=-1}^{3}Im(B_{\frac{m}{2}})Im(HF_{\frac{m}{2}n})+\\
&+\sum_{m=-1}^{5}Re(C_{\frac{m}{2}})Re(HV_{\frac{m}{2}n})+\sum_{m=-1}^{5}Im(C_{\frac{m}{2}})Im(HV_{\frac{m}{2}n})\\
&+\sum_{j=0}^{2N}P_{\frac{j}{2}}Im(HFP_{\frac{j}{2}n})+\sum_{j=0}^{2N}Q_{\frac{j}{2}}Re(HFQ_{\frac{j}{2}n})\\\end{align*}

\begin{align*}
Im(\phi^{*}_1(n))&=-\sum_{m=-1}^{3}Re(B_{\frac{m}{2}})Im(HF_{\frac{m}{2}n})+\sum_{m=-1}^{3}Im(B_{\frac{m}{2}})Re(HF_{\frac{m}{2}n})-\\
&-\sum_{m=-1}^{5}Re(C_{\frac{m}{2}})Im(HV_{\frac{m}{2}n})+\sum_{m=-1}^{5}Im(C_{\frac{m}{2}})Re(HV_{\frac{m}{2}n})-\\
&-\sum_{j=0}^{2N}P_{\frac{j}{2}}Re(HFP_{\frac{j}{2}n})+\sum_{j=0}^{2N}Q_{\frac{j}{2}}Im(HFQ_{\frac{j}{2}n})\\
\end{align*}

\begin{align*}
Re(\phi^{*}_0(n))&=\sum_{m=-1}^{1}Re(A_{\frac{m}{2}})Re(HF_{\frac{m}{2}n})+\sum_{m=-1}^{1}Im(A_{\frac{m}{2}})Im(HF_{\frac{m}{2}n})+\\
&+\sum_{j=0}^{2N}P_{\frac{j}{2}}Re(HF_{\frac{p}{2}n})+\sum_{j=0}^{2N}Q_{\frac{j}{2}}Im(HF_{\frac{p}{2}n})\\
\end{align*}

\begin{align*}
Im(\phi^{*}_0(n))&=\sum_{m=-1}^{1}Im(A_{\frac{m}{2}})Re(HF_{\frac{m}{2}n})-\sum_{m=-1}^{1}Re(A_{\frac{m}{2}})Im(HF_{\frac{m}{2}n})-\\
&-\sum_{j=0}^{2N}P_{\frac{j}{2}}Im(HF_{\frac{j}{2}n})-\sum_{j=0}^{2N}Q_{\frac{j}{2}}Re(HF_{\frac{p}{2}n})\\
\end{align*}

\begin{align*}
h_{ip}&=\binom{p+i+1/2}{p}\left(\frac{L}{r_0}\right)^{p+i+3/2}\\
HF_{np}&=\sum_{i=0}^{\infty}f_{in}h_{ip}\cos(p+i+3/2)\theta_0-i \sum_{i=0}^{\infty}f_{in}h_{ip}\sin(p+i+3/2)\theta_0\\
HV_{np}&=\sum_{i=0}^{\infty}v_{in}h_{ip}\cos(p+i+3/2)\theta_0
-i \sum_{i=0}^{\infty}v_{in}h_{ip}\sin(p+i+3/2)\theta_0\\
HFQ_{np}&= HF_{np}(\alpha L n+2\alpha L)+HF_{(n+1)p}(n\beta L^{3/2}+3/2 \beta L^{3/2})+\\
&+HF_{(n+1)p}(n\gamma L^2 + 4 \gamma L^2) + \delta_{n1}(\alpha L+\beta L^{3/2}+\gamma  L^2)HF_{(n-1)p}\\
HFP_{np}&=n(HF_{np}\alpha L +HF_{(n+1/2)p}\beta L^{3/2} +\\
&+HF_{(n+1)p} \gamma L^2-(\alpha L+\beta L^{3/2}+\gamma L^2)HF_{(n-1)p})
\end{align*}

%\end{widetext}
\footnotesize
\bibliographystyle{elsarticle-num}
\bibliography{enPassant-paper}
\end{document}